\documentclass[aps,prd,groupedaddress,nofootinbib
]{revtex4}
\usepackage{graphicx,amssymb,bm}
\usepackage{subfigure}
\usepackage{microtype}
\usepackage{color}

\newcommand{\be}{\begin{equation}}
\newcommand{\ee}{\end{equation}}
\newcommand{\bea}{\begin{eqnarray}}
\newcommand{\eea}{\end{eqnarray}}
\newcommand{\nab}{\nabla}
\newcommand{\nn}{\nonumber}
\def\rf#1{(\ref{#1})}

\def\prjnab{ \widetilde{\nabla}}

\usepackage{amsmath,amssymb}
\def \curl {\mbox{curl}\,}
\begin{document}

\title{Covariant gauge invariant theory of Scalar Perturbations in $f(R)$-gravity:\\ a brief review.}

\author{Sante Carloni}

\affiliation{Institut d'Estudis Espacials de Catalunya (IEEC) \\
Campus UAB, Facultat Ci\`encies, Torre C5-Par-2a pl \\ 
E-08193 Bellaterra(Barcelona) Spain}
\begin{abstract}
I review the state of the art of the investigation on the structure formation in $f(R)$-gravity based on the Covariant and Gauge Invariant approach to perturbations. A critical analysis of the results, in particular the presence of characteristic signature of these models, together with their meaning and their implication is given.
\end{abstract}
\maketitle

\tolerance=5000

\section{Introduction}
Determining the nature of dark energy \cite{DE} has become, undoubtedly, one of the central issues in modern cosmology as well as in theoretical physics. The interest in this problem has grown to such level that nowadays much of the research work in theoretical cosmology is focused on the quest of a theoretical framework in which dark energy can be explained. One idea that has recently gained much popularity is that dark energy has a geometrical nature i.e. it is originated not by some unknown  new form of energy density that dominate the cosmic evolution, but by the fact that the gravitational interaction, at least on cosmological scales has to be modified. 

Among the many modified versions of General Relativity (GR), higher order theories of gravity, and particularly  one of its subclass, called  $f(R)$-gravity, has been thoroughly studied (see e.g. \cite{Nojiri:2006ri,Capozziello:2007ec,Revnostra} for some reviews). In these models the action for the gravitational interaction is written as a generic analytic function of the Ricci scalar and they can be recovered naturally in the low energy limit of fundamental theories (see e.g. \cite{Nojiri:2006je}). However, the reason why those models have attracted such attention is that it has been proven in many different ways that these theories are able to reproduce in a natural way the cosmological ``footprint" of Dark Energy i.e. the cosmic acceleration \cite{Nojiri:2006ri,Capozziello:2007ec,Sotiriou:2008rp,Carloni:2004kp,SanteGenDynSys}. Such phase cannot be achieved if one considers only GR without the introduction of some additional field with suitable (and, unfortunately, odd) thermodynamical properties.

However, $f(R)$-gravity is not the only model capable to achieve this kind of results and the modification of a theory as well established as GR requires a serious, careful and unbiased  analysis. To be completely fair so far although many concerns have been raised, there is no definite argument against $f(R)$-gravity or, for what matters, an argument that proves definitively the presence of corrections to the gravitational interaction. Therefore there is a great need to devise observable features of these models and compare them to the observations. In this respect cosmology is of great help. In fact, the cosmological models of any theory of the gravitational interaction are by far the easiest application of these schemes and, because of the very nature of the cosmological processes, their analysis is able to give insights of the behavior of these theories in a wide range of energy scales.

Following this spirit in this paper I  will review what it has been discovered so far on the behavior of the scalar (matter) perturbations and  the structure formation in this framework. On this topic many papers have been published that employ a variety of different techniques$/$approximations \cite{otherperts}. In what follows I  will only consider  the results obtained with a specific method: the Covariant Gauge Invariant (CoGI) approach. This approach was proposed at the beginning of the eighties by Bruni and Ellis \cite{EB} and  generalized also by Hwang and Dunsby \cite{EBH,DBE,BED,BDE}. It is constructed on a formalism optimized to analyze cosmological models: the 1+3 covariant approach \cite{EllisCovariant,Ehlers1+3}. Using the CoGI approach we will realize that the behavior of the perturbation in this type of fourth order gravity can be very different from the one of the standard picture, but not necessarily incompatible with the observations. I  will show that the process of structure formation offers in some cases very specific signatures, which can be easily tested with the data currently available.

The paper will be organized as follows. In Section II  the general equations for $f(R)$-gravity are briefly described. In Section III I will review briefly the 1+3 covariant approach to cosmology and its application to $f(R)$-gravity. In Section IV on the base of the 1+3 approach the CoGI approach is developed for the study of the scalar perturbations. In Section V I will summarize the results obtained for some simple specific models. Finally section VI is dedicated to the conclusions.

Unless otherwise specified, natural units
($\hbar=c=k_{B}=8\pi G=1$)
will be used throughout the paper, Latin indices running from 0 to 3.
The symbol $\nabla$ represents the usual covariant derivative and
$\partial$ corresponds to partial differentiation. I  use the
$-,+,+,+$ signature and the Riemann tensor is defined by
\begin{equation}
R^{a}{}_{bcd}=W^a{}_{bd,c}-W^a{}_{bc,d}+ W^e{}_{bd}W^a{}_{ce}-
W^f{}_{bc}W^a{}_{df}\;,
\end{equation}
where the $W^a{}_{bd}$ are the Christoffel symbols (symmetric in
the lower indices), defined by
\begin{equation}
W^a_{bd}=\frac{1}{2}g^{ae}
\left(g_{be,d}+g_{ed,b}-g_{bd,e}\right)\;.
\end{equation}
The Ricci tensor is obtained by contracting the {\em first} and the
{\em third} indices
\begin{equation}\label{Ricci}
R_{ab}=g^{cd}R_{acbd}\;.
\end{equation}
Symmetrization and antisymmetrization  over the indices of a tensor are defined as
\begin{equation}
T_{(a b)}= \frac{1}{2}\left(T_{a b}+T_{b a}\right)\;,\qquad \qquad T_{[a b]}= \frac{1}{2}\left(T_{a b}-T_{b a}\right)\,.
\end{equation}
Finally, the Hilbert--Einstein action in the presence of matter is
given by
\begin{equation}
{\cal A}=\int d x^{4} \sqrt{-g}\left[R+ 2\mathcal{L}_{m}\right]\;.
\end{equation}

\section{General equations for fourth order gravity.}
In four dimensional homogeneous and isotropic spacetimes i.e. Friedmann Lema\^{\i}tre Robertson Walker (FLRW) universes, a general action for fourth order gravity can be written as an analytic function of the Ricci scalar only:
\begin{equation}\label{lagr f(R)}
\mathcal{A}=\int d^4 x \sqrt{-g}\left[ f(R)+2{\cal L}_{m}\right]\;,
\end{equation}
where $\mathcal{L}_m$ represents the matter contribution.
Varying the action with respect to the metric gives the generalization of the Einstein equations:
\begin{equation}\label{eq:einstScTn}
f'G_{ab}=f'\left(R_{ab}-\frac{1}{2}\,g_{ab} R\right)=T
_{ab}^{m}+\frac{1}{2}g_{ab} \left(f-R f'\right) +\nab_b\nab_a f'-
g_{ab}\nab_c\nab^c f'\;,
\end{equation}
where $f=f(R)$, $f'= \displaystyle{\frac{d f(R)}{dr}}$, and
$\displaystyle{T^{M}_{ab}=\frac{2}{\sqrt{-g}}\frac{\delta
(\sqrt{-g}\mathcal{L}_{m})}{\delta g_{ab}}}$ represents the
stress energy tensor of standard matter. These equations reduce to
the standard Einstein field equations when $f(R)=R$. It is crucial
for our purposes to be able to write \rf{eq:einstScTn} in the form\footnote{For this step to make sense it is crucial that we suppose $f'(R)\neq 0$ at all time. This could be problematic when one deals with cosmologies in which the Ricci scalar is zero (like the radiation dominated ones) and functions $f$ such that $f'(0)=0$. In the following, although we will retain a general barotropic factor, we will consider matter to be dust in all but one case.}
\begin{equation}
\label{eq:einstScTneff}
 G_{ab}=\tilde{T}_{ab}^{m}+T^{R}_{ab}=T^{tot}_{ab}\,,
 \end{equation}
where $\displaystyle{\tilde{T}_{ab}^{m}=\frac{ T_{ab}^{m}}{f'}}$ and
\begin{eqnarray}\label{eq:TenergymomentuEff}
T_{ab}^{R}=\frac{1}{f'}\left[\frac{1}{2}g_{ab} \left(f-R f'\right)
+\nab_b\nab_a f'- g_{ab}\nab_c\nab^cf'\right], \label{eq:semt}
\end{eqnarray}
represent two effective ``fluids": the  {\em curvature ``fluid"}
(associated with $T^{R}_{ab}$) and  the {\em effective matter
``fluid"} (associated with $\tilde{T}_{ab}^{m}$) \cite{Revnostra,SantePert}. In this way fourth order gravity can be treated as standard Einstein gravity plus two ``effective" fluids and we can adapt easily many of the techniques developed for GR-based models.
 
Let us look at the conservation properties of these effective fluids using the Bianchi identities \cite{Carloni:2004kp}. The covariant derivative of the total stress energy momentum in \rf{eq:einstScTneff} yields:
\begin{equation}
0=\nabla^bT^{tot}_{ab}=\frac{\nabla^b{T}_{ab}^{m}}{f'}-\frac{f''}{(f')^2} {T}_{ab}^{m} \nabla^bR+\nabla^bT^{R}_{ab}
\end{equation}
Using the field equations and the definition of the Riemann tensor, it is easy to show that the 
sum of the first two terms on the RHS of the previous expression is zero. Thus, $T^{tot~;b}_{~ab}\propto {T}_{ab~;b}^{m} $ and the total conservation equation reduces to the one for matter only. 
A general, independent proof of this result has been given by Eddington \cite{Eddington} and then by others (like in \cite{Lovelock}). They showed that the first variation for the gravitational action is divergence free regardless of the 
form of the invariants that we choose for the Lagrangian. This means that no matter how 
complicated the effective stress--energy tensor $T^{tot}_{~ab}$ is, it will always be divergence free if ${T}_{ab~;b}^{m}=0$. As a consequence, the total conservation equation reduces to the one for standard 
matter only. For our purposes this is important because it tells us that  no matter how the  effective fluids behave, standard matter still follows the usual conservation equations $T_{ab}^{m\ ;b}=0$.

The form \rf{eq:einstScTneff} of the field equations  allows us to use directly the 1+3 covariant approach that will be introduced in the next section.
\section{The  1+3 covariant approach to cosmology }\label{CoVApp}
Basically all the calculations and the results that will be given in this paper are based on the {\it 1+3 covariant approach} \cite{EllisCovariant,Ehlers1+3}. Such approach is, conceptually, not different from the ADM one, but it is specifically adapted to treat homogeneous and isotropic spacetimes and can simplify quite a lot the calculations. Therefore, before starting, it is worth to give a very brief review of this approach and its application to $f(R)$-gravity.

The 1+3 approach is based on the choice of a specific family of preferred worldlines which represent specific classes of observers and  can be associated to a timelike vector field $u^a$. Using this field  one can split the metric tensor as
\be\label{gdecom1+3}
g_{ab} = h_{ab} - u_a\,u_b\:,
\ee
i.e. the spacetime is foliated in hypersurfaces with metric $h_{ab}$ orthogonal to the vector field $u_{a}$. In this way any affine parameter on the worldlines associated to  $u_a$ can be chosen to represent ``time" and the tensor $h_{ab}$  ($h^a{}_c\,h^c{}_b = h^a{}_b \ ,~ h^a{}_a = 3 \ , ~h_{ab}\,u^b = 0$) determines the geometry of the instantaneous rest-spaces of the observers we have chosen.  Using $u_a$ and $h_{ab}$, one can then define the projected volume form $\eta_{abc}=u^d\eta_{abcd}$ on the rest spaces, the covariant time derivative ($\,\dot{}\,$) along the fundamental worldlines, and the  fully orthogonally projected covariant derivative $\prjnab$:
\be
\dot{X}^{ab}{}_{cd} = u^{e}\nabla_{e}X^{ab}{}_{cd} \ , \qquad
\prjnab_{e}X^{ab}{}_{cd} = h^a{}_f\,h^b{}_g\,h^p{}_c\,h^q{}_d\,
h^r{}_e\nabla_r\,X^{fg}{}_{pq} \ .
\ee
Also, performing a split of the first covariant derivative of $u_a$ into its irreducible parts, namely
\be
\label{eq:kin}
\nabla_{a}u_{b} =
-\,u_a\,a_b + {\frac{1}{3}}\,\Theta\,h_{ab} + \sigma_{ab}
+ \omega_{ab} \, ,
\ee
one can define, in analogy with classical hydrodynamics \cite{Ehlers1+3, Plebanski:2006sd}, the basic kinematical quantities of this formalism \cite{EllisCovariant}.
In \rf{eq:kin}, $\Theta = \prjnab_au^a{}$ represents the rate
of volume expansion of the worldlines of $u_a$  so that the standard Hubble parameter is $H$: $H=3\Theta$; $\sigma_{ab} = \prjnab_{\langle a}u_{b\rangle}$
is the trace-free symmetric  rate of shear tensor
 describing the rate of distortion of the
observer flow and we have $\sigma_{ab} = \sigma_{(ab)}$, $\sigma_{ab}\,u^b = 0$,
$\sigma^a{}_{a}= 0$; $\omega_{ab} = \prjnab_{[a}u_{b]}$ is the
skew-symmetric vorticity tensor describing the rotation of the observers relative 
to a Fermi-propagated (non-rotating) frame and $\omega_{ab} =
\omega_{[ab]}$, $\omega_{ab}\,u^b = 0$; $a_b=\dot{u}_b$ is the acceleration vector,
which describes the non-gravitational forces acting on the observers.

The matter energy-momentum tensor $T_{ab}$ of a general fluid can also be
decomposed locally using $u_a$ and $h_{ab}$:
\bea
T_{ab} = \mu\,u_a\,u_b + q_a\,u_b + u_a\,q_b + p\,h_{ab} + \pi_{ab}\;,
\label{eq:stress}
\eea
where $\mu = (T_{ab}u^{a}u^{b})$ is the relativistic energy
density relative to $u^a$, $q^{a} = -\,T_{bc}\,u^{b}\,h^{ca}$  is
the  relativistic momentum density, which is also the energy
flux relative to $u^a$ ($q_a\,u^a = 0 $), $p = {\frac{1}{3}}\,(T_{ab}h^{ab})$ is the
 isotropic pressure, and $\pi_{ab} = T_{cd}\,h^{c}{}_{\langle
a}\,h^{d}{}_{b\rangle}$ 
is the trace-free  anisotropic pressure for which $\pi^a{}_a = 0 \ , ~\pi_{ab} = \pi_{(ab)} $. This allows us to extract information 
on the thermodynamics associated to this fluid.

The kinematics and thermodynamics quantities presented above completely determine a cosmological model. The advantages in using these variables is that they allow a treatment of cosmology that is both mathematically rigorous and  physically meaningful  and they are particularly useful in the construction of the theory of perturbations. Their evolution and constraint equations, also known as {\it 1+3 covariant equations} (see \cite{EllisCovariant}), are completely equivalent to the Einstein equations and characterize  the full evolution of the cosmology. 
\subsection{The 1+3 covariant approach for $f(R)$-gravity}
Following the above scheme let us apply the 1+3 formalism to $f(R)$-gravity.  As first step one needs to choose suitable frame, i.e., a 4-velocity field $u_{a}$. Following \cite{SantePert,StructForm}, we will choose the frame $u^{m}_a$ comoving with standard matter represented by galaxies and clusters of galaxies. This frame basically coincides to our specific point of view: as earth bound observers we are comoving with these object. We will also
assume that in $u^{m}_a$ standard matter is a barotropic perfect fluid with equation of state $p^m=w \mu^m$. 

Relative to $u_a^m$, the stress energy tensor $T^{tot}_{ab}$ given in \rf{eq:einstScTneff} can be decomposed as
\begin{eqnarray}\label{mutot}
\mu^{\rm tot}\,&=&T^{\rm tot}_{ab}u^{a}u^{b}\,=\,\tilde{\mu}^{\,
m}+\mu^{\,R}\,,\qquad
p^{\rm tot}\,=\frac{1}{3}T^{\rm tot}_{ab}h^{ab}\,=\,\tilde{p}^{\, m}+p^{\,R}\;,\\
q^{\rm tot}_{a}\,&=&-T^{\rm
tot}_{bc}h_{a}^{b}u^{c}\,=\,\tilde{q}^{\,
m}_{a}+q^{\,R}_{a}\,,\qquad \pi^{\rm tot}_{ab}\,=\,T^{\rm
tot}_{cd}h_{<a}^{c}h_{b>}^{d}\,=\,\tilde{\pi}^{\,
m}_{ab}+\pi^{\,R}_{ab}\,,
\end{eqnarray}
with
\begin{eqnarray}
\tilde{\mu}^{\,m}\,&=&\,\frac{\mu^{\,m}}{f'}\,,\qquad
\tilde{p}^{\,m}\,=\,\frac{p^{\,m}}{f'}\,,\qquad
\tilde{q}^{\,m}_{a}\,=\,\frac{q^{\,m}_{a}}{f'}\,,\qquad
\tilde{\pi}^{\,m}_{ab}=\,\frac{\pi^{\,m}_{ab}}{f'}\, .
\end{eqnarray}
Since we assume that  standard matter is a perfect fluid in $u_a^m$,
$q^{\,m}_{a}$ and $\pi^{\,m}_{ab}$ are zero, so that the last two
quantities above  also vanish. The effective thermodynamical quantities for the curvature ``fluid"
are
\begin{eqnarray}
&&\mu^{R}\,=\,\frac{1}{f'}\left[\frac{1}{2}(R f'-f)-\Theta
f''\dot{R}+f''\tilde{\nabla}^2{R}+f'''\,\prjnab^b{R}\prjnab_b{R}\right]\;,\label{muR}\\
&&p^{R}\,=\,\frac{1}{f'}\left[\frac{1}{2}(f-R
f')+f''\ddot{R}+f'''\dot{R}^2+\frac{2}{3}\Theta
f''\dot{R}-\frac{2}{3}f''\tilde{\nabla}^2{R} +\right. \nonumber\\
&& \qquad\left.  -\frac{2}{3}f'''\prjnab^{a}{R}\prjnab_{a}{R}+f''
\,a_b\prjnab^b{R}\right]\;,\label{pR}\\
&&q^{R}_a\,=\,-\frac{1}{f'}\left[f'''\dot{R}\prjnab_{a}R+f''\prjnab_{a}\dot{R}-\frac{1}{3}\Theta f''
\prjnab_{a}R\right]\;,\\
&&\pi^{R}_{ab}\,=\,\frac{1}{f'}\left[f''\prjnab_{\langle
a}\prjnab_{b\rangle}R+f'''\prjnab_{\langle a}{R}\prjnab_{b\rangle}{R}-\sigma_{a
b}f''\dot{R}\right]\,.\label{piR}
\end{eqnarray}
The twice-contracted Bianchi Identities lead to evolution equations for
$\mu^{\,m}$, $\mu^{R}$, $q^{R}_a$ and are given in \cite{SantePert}. 
In this way the 1+3 equations for $f(R)$-gravity in the frame $u_a^m$ can be written as:
\\\\
\noindent Expansion propagation (generalized Raychaudhuri equation):
\begin{eqnarray}\label{1+3eqRayHO}
&&\dot{\Theta}+{\textstyle\frac{1}{3}}\Theta^2+\sigma_{{{a}}{{b}}}
\sigma^{{{a}}{{b}}} -2\omega_{{a}}\omega^{{a}} -\tilde{\nabla}^a
\dot{u}_{{a}}+ \dot{u}_{{a}}
\dot{u}^{{a}}+{\textstyle\frac{1}{2}}(\tilde{\mu}^{m} +
3\tilde{p}^{m}) =-{\textstyle\frac{1}{2}}({\mu}^{R} +
3{p}^{R})\;.
\end{eqnarray}
Vorticity propagation:
\begin{equation}\label{1+3VorHO}
\dot{\omega}_{\langle {{a}}\rangle }
+{\textstyle\frac{2}{3}}\Theta\omega_{{a}} +{\textstyle\frac{1}{2}}\curl
\dot{u}_{{a}} -\sigma_{{{a}}{{b}}}\omega^{{b}}=0 \;.
\end{equation}
Shear propagation:
\begin{equation}\label{1+3ShearHO}
\dot{\sigma}_{\langle {{a}}{{b}} \rangle }
+{\textstyle\frac{2}{3}}\Theta\sigma_{{{a}}{{b}}}
+E_{{{a}}{{b}}}-\prjnab_{\langle {{a}}}\dot{u}_{{{b}}\rangle }
+\sigma_{{c}\langle {{a}}}\sigma_{{{b}}\rangle }{}^{c}+
\omega_{\langle {{a}}}\omega_{{{b}}\rangle} - \dot{u}_{\langle
{{a}}}\dot{u}_{{{b}}\rangle}
\,=\,{\textstyle\frac{1}{2}}\pi^{R}_{{{a}}{{b}}}\;.
\end{equation}
Gravito-electric propagation:
\begin{eqnarray}\label{1+3GrElHO}
 && \dot{E}_{\langle {{a}}{{b}} \rangle }
+\Theta E_{{{a}}{{b}}} -\curl H_{{{a}}{{b}}}
+{\textstyle\frac{1}{2}}(\tilde{\mu}^{m}+\tilde{p}^{m})\sigma_{{{a}}{{b}}}
-2\dot{u}^{c}\eta_{{c}{d}({{a}}}H_{{{b}})}{}^{d} -3\sigma_{{c}\langle
{{a}}}E_{{{b}}\rangle }{}^{c} +\omega^{c}
\eta_{{c}{d}({{a}}}E_{{{b}})}{}^{d}
\nonumber\\&&~~{}=-{\textstyle\frac{1}{2}}(\mu^{R}+p^{R})\sigma_{{{a}}{{b}}}
-{\textstyle\frac{1}{2}}\dot{\pi}^{R}_{\langle {{a}}{{b}}\rangle  }
-{\textstyle\frac{1}{2}}\prjnab_{\langle {{a}}}q^{R}_{{{b}}\rangle }
-{\textstyle\frac{1}{6}}
\Theta\pi^{R}_{{{a}}{{b}}}-{\textstyle\frac{1}{2}}\sigma^{c}{}_{\langle
{{a}}}\pi^{R}_{{{b}}\rangle {c}} -{\textstyle\frac{1}{2}}
\omega^{c}\eta_{{c}({{a}}}^{d}\pi^{R}_{b )d}\;.
\end{eqnarray}
Gravito-magnetic propagation:
\begin{eqnarray}\label{1+3GrMagHO}
 &&\dot{H}_{\langle
{{a}}{{b}} \rangle } +\Theta H_{{{a}}{{b}}} +\curl E_{{{a}}{{b}}}-
3\sigma_{{c}\langle {{a}}}H_{{{b}}\rangle }{}^{c} +\omega^{c}
\eta_{{c}{d}({{a}}}H_{{{b}})}{}^{d}
+2\dot{u}^{c}\eta_{{c}{d}({{a}}}E_{{{b}})}{}^{d}
\nonumber\\&&~~{}={\textstyle\frac{1}{2}}\curl\pi^{R}_{{{a}}{{b}}}-{\textstyle\frac{3}{2}}\omega_{\langle
{{a}}}q^{R}_{{{b}}\rangle
}+{\textstyle\frac{1}{2}}\sigma^{c}{}_{({{a}}}
\eta_{{{b}}){c}}^{\;\;\;\;d}q^{R}_{d}\;.
\end{eqnarray}
Vorticity constraint:
\begin{equation}\label{1+3VorConstrHO}
\prjnab^{{a}}\omega_{{a}} -\dot{u}^{{a}}\omega_{{a}} =0\;.
\end{equation}
Shear constraint:
\begin{equation}\label{1+3ShearConstrHO}
\prjnab^{{b}}\sigma_{{{a}}{{b}}}-\curl\omega_{{a}}
-{\textstyle\frac{2}{3}}\prjnab_{{a}}\Theta +2[\omega,\dot{u}]_{{a}} =
-q^{R}_{a}\;.
\end{equation}
Gravito-magnetic constraint:
\begin{equation}\label{1+3GrMagConstrHO}
 \curl\sigma_{{{a}}{{b}}}+\prjnab_{\langle {{a}}}\omega_{{{b}}\rangle  }
 -H_{{{a}}{{b}}}+2\dot{u}_{\langle {{a}}}
\omega_{{{b}}\rangle  }=0 \;.
\end{equation}
Gravito-electric divergence:
\begin{eqnarray}\label{1+3GrElConstrHO}
&& \prjnab^{{b}} E_{{{a}}{{b}}}
-{\textstyle\frac{1}{3}}\prjnab_{{a}}\tilde{\mu}^{m} -[\sigma,H]_{{a}}
+3H_{{{a}}{{b}}}\omega^{{b}}={\textstyle\frac{1}{2}}\sigma_{{{a}}}^{{b}}q^{R}_{{b}}-
{\textstyle\frac{3}{2}}
[\omega,q^{R}]_{{a}}-{\textstyle\frac{1}{2}}\prjnab^{{b}}\pi^{R}_{{{a}}{{b}}}
 +{\textstyle\frac{1}{3}}\prjnab_{{a}}\mu^{R}
-{\textstyle\frac{1}{3}}\Theta q^{R}_{{a}}\;.
\end{eqnarray}
Gravito-magnetic divergence:
\begin{eqnarray}\label{1+3GrMagDivHO}
 &&\prjnab^{{b}} H_{{{a}}{{b}}}
-(\tilde{\mu}^{m}+\tilde{p}^{m})\omega_{{a}} +[\sigma,E]_{{a}}
 -3E_{{{a}}{{b}}}\omega^{{b}}=-{\textstyle\frac{1}{2}}\curl q^{R}_{{a}}
+(\mu^{R}+p^{R})\omega_{{a}} -{\textstyle\frac{1}{2}}
[\sigma,\pi^{R}]_{{a}} -{\textstyle\frac{1}{2}}\pi^{R}_{{{a}}{{b}}}
\omega^{{b}}\;.
\end{eqnarray}
Standard Matter Conservation (twice contracted Bianchi identities)
\begin{eqnarray}
&&\dot{\mu}^m\,=\, - \,\Theta\,(\mu^m+{p^m})\;,\label{eq:cons1}\\
&&\prjnab^{a}{p^m} =  - (\mu^m+{p^m})\,\dot{u}^{a}\,.
\end{eqnarray}
Curvature fluid Conservation (twice contracted Bianchi identities)
\begin{eqnarray}
&&\label{eq:cons2} \dot{\mu^R} + \prjnab^{a}q^R_{a} = - \,\Theta\,(\mu^R+p^R)
- 2\,(\dot{u}^{a}q^R_{a}) -
(\sigma^{a}\!^{b}\pi^R_{b}\!_{a})+\mu^{m}\frac{f''\,\dot{R}}{f'^{2}}\;,\\
&& \label{eq:cons3} \dot{q}^R_{\langle a\rangle} + \prjnab_{a}p^R +
\prjnab^{b}\pi^R_{ab} = - \,{\textstyle\frac{4}{3}}\,\Theta\,q^R_{a} -
\sigma_{a}\!^{b}\,q^R_{b} - (\mu^R+p^R)\,\dot{u}_{a} -
\dot{u}^{b}\,\pi^R_{ab} -
\eta_{a}^{bc}\,\omega_{b}\,q^R_{c}+\mu^{m}\frac{f''\,\prjnab_{a}{R}}{f'^{2}}
\  \end{eqnarray}
In the equations above the spatial curl of a vector and a tensor is
\begin{equation}
(\curl\,X)^{a} = \eta^{abc}\,\prjnab_{b}X_{c}\,,\qquad \qquad (\curl\,X)^{ab} = \eta^{cd\langle a}\,\prjnab_{c}X^{b\rangle}\!_{d}\,,
\end{equation}
respectively.

Finally, $\omega_{{a}}=\frac{1}{2}\eta_{a}{}^{{b}{c}}\omega_{bc}$ and the
covariant  commutators are
\[
[X,Y]_{{a}}=\eta_{{{a}}{c}{d}}X^cY^d\;,\qquad [W,Z]_{{a}} =\eta_{{{a}}{c}{d}}W^{c}{}_{e} Z^{{d}{e}}\,.
\]
Note that for $f(R)=R$, one has $f'(R)=1$ and $\mu^R, p^R, q_a^R, \pi_{a b}^R=0$ and therefore the above equations reduce to the ones for GR. Using \rf{1+3eqRayHO}-\rf{eq:cons3} one can analyze any type of $f(R)$ cosmology. Such analysis can be performed either directly or with the use of alternative techniques, like the Dynamical System Approach (see e.g. \cite{Carloni:2004kp,SanteGenDynSys,shosho,Leach:2006br}).

\section{Covariant gauge Invariant Perturbations}
Once the evolution of a cosmological model has been determined, one can investigate the perturbations around a specific metric which is a solution for the background cosmology\footnote{It is worth to stress here that this last prescription is a particularly important one. Choosing a random background which is incompatible with the 0 order cosmological equations can potentially destroy the entire predictive power of the perturbation equations.}. In order to analyze the behavior of the perturbations we will  use the 1+3 covariant approach described above to construct a Covariant Gauge Invariant (CoGI) theory of perturbations. Such theory will characterize the evolution of the perturbations as covariant equations written in terms of variables which have a very straightforward physical meaning  and are naturally gauge invariant.

The first step is the choice of the background. In the CoGI approach this is not done by assigning a metric, but rather by recognizing which 1+3 quantities are zero in the background and which are not. In what follows we will consider expanding ($\Theta\neq0$) homogeneous and isotropic ($\sigma_{ab}=0$, $\omega_{ab}=0$) backgrounds. In this setting we will characterize the perturbations in terms the of 1+3 quantities seen in the previous Section and their projected gradients.

For example, to characterize the energy density perturbation, the natural variable
is $X_a=\prjnab_a\mu$. This vector represents any spatial
variation of the energy density $\mu$ (i.e. any over-density or
void) and it is in principle directly measurable \cite{EB}.
However, a more suitable quantity to describe density perturbation
is
\begin{equation}\label{eq:vardefD}
D_a=\frac{ S }{\mu}X_a =\frac{ S }{\mu}\prjnab_a\mu\;,
\end{equation}
where the ratio $X_a/\mu$ allows one to evaluate the magnitude of
density perturbations relative to the background energy density and
the presence of the scale factor $S$ guarantees that it is
dimensionless and comoving in character. The magnitude of $D_a$  ($D=\sqrt{D_aD^a}$) is
closely related with the quantity
$\delta\mu/\mu$, but represents a real  covariant and gauge
invariant spatial fluctuation. In fact, it is easy to
show that the Bardeen variable $\epsilon_m$ which corresponds to
$\delta\mu/\mu$ in the comoving  gauge, is
the scalar harmonic component of $D_a$ (i.e. its scalar
``potential")\cite{BDE}.

Other important quantities relevant to the evolution of density
perturbations  in GR are
\begin{equation}
 Z_a \equiv  S
\prjnab_a\Theta\;,
 ~~ C_a \equiv   S^3 \prjnab_a{R}_{3}\;, \label{eq:vardefZC}
\end{equation}
which represent, the spatial gradient of the expansion and the
spatial gradient of the 3-Ricci scalar respectively. These variables 
are not independent from (\ref{eq:vardefD}), but they are related by a constraint
coming from the  spatial derivative of the Gauss equation \cite{SantePert,StructForm}.
Moreover it can be proven that these variables, as well as any other quantity which vanish in the background, are gauge-invariant \cite{bi:stewart}.

The variables defined above contain a lot of information, such as evolution of non-spherically symmetric structures and vortical motions, which is not directly associated to the evolution of matter fluctuations. In the following we will focus on this last type of perturbations only and, specifically, on  the ones directly related to spherically symmetric collapse. To extract this information from the variables \rf{eq:vardefD} and \rf{eq:vardefZC}, we use the local splitting 
\begin{equation}\label{localdecomposition}
\prjnab_{a}X_b=X_{ab}= \frac{1}{3}h_{ab}X+\Sigma^{X}_{ab}+X_{[ab]}\;,
\end{equation}  where
\begin{equation}
X=\prjnab_aX^{a}\;,
\end{equation}
and
\begin{equation}
\Sigma^{X}_{ab}=X_{(ab)}-\frac{1}{3}h_{ab}X\;.
\end{equation}
Because of its geometrical properties it is clear that only the scalar part $X$ can describe a spherically symmetric collapse. As a consequence, from now on, we will deal only with the scalar part of the perturbation variables.
Such part can be extracted applying the comoving differential operator $S\prjnab_a$ to \rf{eq:vardefD} and \rf{eq:vardefZC}:
\begin{equation}\label{ScaVar}
\Delta_{m}=\frac{S^2}{\mu^{m}}\prjnab^2\mu^{m}\,,\qquad
Z=S^2\prjnab^2\Theta\,,\qquad C=S^{4}\prjnab^2{R}_{3}\;.
\end{equation}
Clearly these variables are gauge invariant, for the same reasons for which $D_a, Z_a, C_a$ are.  In dealing with $f(R)$-gravity, however, \rf{ScaVar} will not be able to describe all the degrees of freedom of the theory. In the next section we will see how this can be addressed.

\subsection{CoGI Perturbations for $f(R)$-gravity}
Let us now derive explicitly the perturbation equations  around an homogeneous and isotropic background for $f(R)$-gravity as seen by an observer associated to $u_a^m$  and in presence of a fluid which is perfect in this frame and has an equation of state $p^m=w \mu^m$.  In this setting the \rf{1+3eqRayHO}-\rf{eq:cons3} give the following background equations:
\begin{subequations}\label{backgroundEQ}
\begin{flalign}
&\Theta^2\,=\,3\frac{\mu^{m}}{f'} + 3\mu^{R}-\frac{3R_3}{2}\;,\\
&\dot{\Theta}+{\textstyle\frac{1}{3}}\Theta^2
+\frac{1}{2 f'}({\mu}^{m} + 3{p}^{m})
        +{\textstyle\frac{1}{2}}({\mu}^{R} + 3{p}^{R})=0\;,\\
& \dot{\mu}^m\,+ \,\Theta\,(\mu^m+{p^m})=0\;,\\
& \dot{\mu}^R\,+ \,\Theta\,(\mu^R+{p^R})-\mu^m\frac{f''}{(f')^2}\dot{R}=0\;,
\end{flalign}
\end{subequations}
where $\mu^{R}$ and $p^{R}$ are given in \rf{muR} and \rf{pR} and $R_3=6K/S^2$ with the spatial curvature index $K=0,\pm1$.

As mentioned above $f(R)$-gravity  contains more degrees of freedom than standard GR. Therefore in order to complete the description of the evolution of the perturbations in this framework  we will need to define some additional perturbation variables. Following the tradition to consider the Ricci scalar an additional effective field of the theory \cite{RField} a choice for these additional variables is given by 
\begin{equation}
 \mathcal{R}_a \equiv  S
\prjnab_a\Theta\;,
 ~~ \Re_a \equiv   S \prjnab_a \dot{R}\;, \label{eq:vardefFR}
\end{equation}
where  ${\mathcal R}_a$  determines the fluctuations in the Ricci scalar  $R$  and  $\Re_a$  the ones of its momentum $\dot{R}$. As for the other perturbation variables here we will consider only their scalar part
\begin{equation}\label{PertVarfR}
{\cal R}=S^2\prjnab^2 R\,,\qquad\Re=S^2\prjnab^2 \dot{R}\;.
\end{equation}
The results of \cite{bi:stewart} will guarantee that these new quantities are indeed gauge invariant. The set of variables $\Delta, Z, C, {\cal R},\Re$  completely characterizes the evolution of  the density perturbations in $f(R)$-gravity once the background is given.

Using equations \rf{backgroundEQ} one can derive the evolution and constraint equations for the variables \rf{ScaVar} and \rf{PertVarfR}. These  equations constitute a system of  first order partial differential equations \cite{SantePert,StructForm}:
\begin{subequations}\label{PertSca}
\begin{flalign}
\dot{\Delta}_m &=w\Theta \Delta_m-(1+w)Z\,,\label{eqDelta}\\
\nn\dot{Z} &= \left(\frac{\dot{R}
   f''}{f'}-\frac{2 \Theta }{3}\right)Z+
   \left[\frac{ (w -1) (3 w +2)}{2 (w +1)} \frac{\mu}{ f'} + \frac{2 w \Theta ^2
   +3 w (\mu^{R}+3  p^{R}) }{6 (w +1) }\right]
   \Delta_m+\frac{\Theta f''}{f'}\Re+\\&\quad+
   \left[\frac{1}{2}-\frac{1}{2} \frac{f}{f'}\frac{ f''}{f'}- \frac{f''}{f'} \frac{\mu}{
   f'} + \dot{R} \Theta  \left(\frac{f''}{f'}\right)^{2}+ \dot{R} \Theta \frac{ f^{(3)}}{ f'}\right]\mathcal{R}
   -\frac{w}{w +1} \prjnab^{2}{\Delta}_m-\frac{ f''}{f'}\prjnab^{2}\mathcal{R}\,,\\
\dot{{\cal R}}&=\Re-\frac{w }{w +1}\dot{R}\;{\Delta}_m\,,\\
\nn\dot{\Re}&=- \left(\Theta + 2\dot{R} \frac{
   f^{(3)}}{f''}\right)\Re- \dot{R} Z -
   \left[\frac{ (3 w -1)}{3} \frac{\mu}{f''} + \frac{w}{3(w +1)} \ddot{R} \right]{\Delta}_m+\\&\quad-\left[\frac{1}{3}\frac{f'}{f''}+\frac{f^{(4)}}{f'} \dot{R}^2+\Theta \dot{R} \frac{f^{(3)}}{f''}+\ddot{R} \frac{f^{(3)}}{f''}-\frac{R}{3}\right]\mathcal{R}+\prjnab^{2}\mathcal{R}\,,\label{eqRho}\\
\dot{C}&=\nonumber k^2 \left[\frac{18  f'' \mathcal{R}}{S^2
\Theta f'}-\frac{18\Delta_{m}}{S^2  \Theta } \right] +K\left[\frac{3}{S^2\Theta}C
+\Delta_{m}\left(\frac{2 (w-1) \Theta }{w +1}+\frac{ 6\mu^{R}}{\Theta}\right)-\frac{6  f''}{\Theta f''}\prjnab^{2}\mathcal{R}
\right.\\\nn&
\quad\left.+\frac{6 f''}{f'}\Re+\frac{ 6 \dot{R} \Theta  f'
   f^{(3)}- f'' \left(3 f-2 \left(\Theta ^2-3 \mu^{R}\right) f'+6 \dot{R} \Theta f''\right)}{\Theta  (f')^2 }\mathcal{R}\right] \\ &\quad+\prjnab^{2}\left[\frac{4 w
    S^2 \Theta }{3 (w +1)}\Delta_{m}+\frac{2  S^2
   f''}{f'}\Re-\frac{2 S^2 \left(\Theta
    f''-3 \dot{R} f^{(3)}\right)}{3
    f'}\mathcal{R}\right]\,,\label{eqC}
\end{flalign}
\end{subequations}
together with the constraint
\begin{equation}\label{Gauss1}
  \frac{C}{S^2}+ \left(\frac{4  }{3}\Theta +\frac{2 \dot{R}
   f''}{f'}\right) Z-2\frac{
    \mu^m }{f'}{\Delta_{m}}+ \left[2 \dot{R}
   \Theta  \frac{f^{(3)}}{ f'}-\frac{f''}{ (f')^2} \left(f-2 \mu^m+2
   \dot{R} \Theta  f''\right)\right]\mathcal{R}+\frac{2 \Theta
   f''}{f'}\Re-\frac{2 f''}{f'}\prjnab^{2}\mathcal{R}=0\,.
\end{equation}
where we have assumed $f''(R)\neq0$ i.e. we are excluding the GR case. Note that this system is made up of four equations, which means that the evolution of every single perturbation variable is determined by a fourth order differential equation. This differs from the case of GR, in which these equations are of second order, and it means that in general the solutions will have four modes. Hence the perturbations in these theories have a much richer structure with respect to the GR ones. In addition to that, the coefficients depend on derivatives of the scale factor of order up to four, so that the behavior of the perturbation is much more sensitive to the feature of the background than GR.
 
In order to reduce the system above to a set of ordinary differential equations, one defines the eigenfunctions of the spatial Laplace-Beltrami operator:
\begin{equation}\label{CovHarmDef}
\prjnab^{2}Q = -\frac{k^{2}}{S^{2}}Q\,,
\end{equation}
where $k=2\pi S/\lambda$ is the wave number and $\dot{Q}=0$,  and expands every first order quantity in the above equations:
\begin{equation}\label{eq:developmentdelta}
X(t,\mathbf{x})=\sum X^{(k)}(t)\;Q^{(k)}(\mathbf{x})\;,
\end{equation}
where $\sum$ stands for both summation over discrete or integration over continuous indices. In this way, one obtains the equations describing the $k^{th}$ mode for scalar perturbations in $f(R)$ gravity. They are \cite{SantePert,StructForm}:
\begin{subequations}\label{SysIOrdHarm}
\begin{flalign}
&\dot{\Delta}_{m}^{(k)} =w\Theta \Delta_{m}^{(k)}-(1+w)Z^{(k)}\,,\label{eqDeltaHarm}\\
&\dot{Z}^{(k)} = \left(\frac{\dot{R}
   f''}{f'}-\frac{2 \Theta }{3}\right)Z^{(k)}+
    \left[\frac{ (w -1) (3 w +2)}{2 (w +1)} \frac{\mu^{m}}{ f'} + \frac{2 w \Theta ^2
   +3 w (\mu^{R}+3  p^{R}) }{6 (w +1) }\right]   \Delta_{m}^{(k)}+\frac{\Theta f''}{f'}\Re^{(k)}+\nonumber \\  &\qquad+
   \left[\frac{1}{2}-\frac{ f''}{f'} \frac{k^2}{S^2}-\frac{1}{2} \frac{f}{f'}\frac{ f''}{f'}- \frac{f''}{f'} \frac{\mu^{m}}{
   f'} + \dot{R} \Theta  \left(\frac{f''}{f'}\right)^{2}+ \dot{R} \Theta \frac{ f^{(3)}}{ f'}\right]\mathcal{R}^{(k)}\,,\\
&\dot{{\cal R}}^{(k)}=\Re^{(k)}-\frac{w }{w +1}\dot{R}\;{\Delta}_{m}^{(k)}\,,\label{eqZHarm}\\
&\dot{\Re}^{(k)}=- \left(\Theta + 2\dot{R} \frac{f^{(3)}}{f''}\right)\Re^{(k)}- \dot{R} Z^{(k)} -
  \left[\frac{ (3 w -1)}{3} \frac{\mu^{m}}{f''} + \frac{w}{3(w +1)} \ddot{R} \right]{\Delta}_{m}^{(k)}+\nonumber\\ & \qquad +\left[\frac{k^{2}}{S^2}-\left(\frac{1}{3}\frac{f'}{f''}+\frac{f^{(4)}}{f'} \dot{R}^2+\Theta \dot{R} \frac{f^{(3)}}{f''}+\ddot{R} \frac{f^{(3)}}{f''}-\frac{R}{3}\right)\right]\mathcal{R}^{(k)}\label{eqRho2}\,,\\
&\dot{C}^{(k)}=\nn K^2 \left[\frac{18  f'' \mathcal{R}}{S^2
\Theta f'}-\frac{18\Delta_{m}}{S^2  \Theta } -6\frac{f''}{\Theta f'}\Re^{(k)}\right]  +K\left[\frac{3}{S^2\Theta}C
+\Delta\left(\frac{2 (w-1) \Theta }{w +1}-\frac{ 6\mu^{R}}{\Theta}\right)-\frac{6  f''}{\Theta f''}\prjnab^{2}\mathcal{R}
\right.\\& \qquad
\nonumber\left.+\frac{6 f''}{f'}\Re +\frac{ 6 \dot{R} \Theta  f'
   f^{(3)}-6 k^2 f'' f'+ f'' \left(3 f-2 \left(\Theta ^2-3 \mu^{R}\right) f'+6 \dot{R} \Theta f''\right)}{\Theta  (f')^2 }\mathcal{R}\right]+\nn\\& \qquad+\frac{k}{S^{2}}\left[\frac{4 w S^2 \Theta }{3 (w +1)}\Delta_{m}^{(k)}+\frac{2  S^2
   f''}{f'}\Re^{(k)}-\frac{2 S^2 \left(\Theta f''-3 \dot{R} f^{(3)}\right)}{3 f'}\mathcal{R}^{(k)}\right]\,,\label{eqCHarm}
\end{flalign}
\end{subequations}
\begin{eqnarray}
 &&\frac{C^{(k)}}{S^2}+ \left(\frac{4  }{3}\Theta +\frac{2\dot{R} f''}{f'}\right) Z^{(k)}-2\frac{
    \mu }{f'}{\Delta}_m^{(k)}+\frac{2 \Theta
   f''}{f'}\Re^{(k)}\nn\\ &&\quad + \left[2 \dot{R}
   \Theta  \frac{f^{(3)}}{ f'}-\frac{f''}{ (f')^2} \left(f-2 \mu +2
   \dot{R} \Theta  f''\right)+2 \frac{ f''}{f'} \frac{k^{2}}{S^2}\right]\mathcal{R}^{(k)}=0\;,\label{constrCZDel}
\end{eqnarray}

This system takes a more manageable form if we reduce it to a pair of second order equations:
\begin{subequations}\label{EqPerIIOrd}
\begin{flalign}
  &\ddot{\Delta}_{m}^{(k)}+ \mathcal{A} \dot{\Delta}_{m}^{(k)}+\mathcal{B}\Delta_{m}^{(k)}=\mathcal{C} \mathcal{R}^{(k)}+\mathcal{D} \dot{\mathcal{R}}^{(k)}\label{EqPerIIOrd1}\,,\\&
  f''\ddot{\mathcal{R}}^{(k)}+\mathcal{E}
  \dot{\mathcal{R}}^{(k)}+\mathcal{F}\mathcal{R}^{(k)}= \mathcal{G}\Delta_{m}^{(k)}+\mathcal{H} \dot{\Delta}_{m}^{(k)}\,.\label{EqPerIIOrd2}
\end{flalign}
\end{subequations}
where
\begin{subequations}\label{CoeffPertEinf'rame}
\begin{flalign}
&\mathcal{A}=\left(\frac{2}{3}-w\right) \Theta -\frac{\dot{R}f''}{f'}\,,\\
&\mathcal{B}=-\left[w  \frac{k^2}{S^{2}}-w  (3 p^{R}+\mu^{R})-\frac{2 w  \dot{R} \Theta
  f''}{f'}-\frac{\left(3 w ^2-1\right) \mu^{m} }{f'}\right],\\
& \mathcal{C}=  \frac{1}{2}(w +1)\left[-2 \frac{k^2}{S^2}\frac{f''}{f'}-1+ \left(f-2 \mu^{m} +2 \dot{R} \Theta  f''\right)\frac{f''}{f'^2}-2  \dot{R} \Theta\frac{f^{(3)}}{f'}\right],\\
&\mathcal{D}= -\frac{(w +1) \Theta f'' }{f'}\,,\\
&\mathcal{E}=\left(\Theta f'' +2 \dot{R} f^{(3)}\right)\,,\\ 
&\nn\mathcal{F}=-\left[\frac{k^2}{S^2}f''+ 2 \frac{K}{S^2}f''
 +\frac{2}{9} \Theta^2 f''- (w +1) \frac{\mu^{m}}{2 f'}f''- \frac{1}{6}(\mu^{R}+ 3
p^{R})f''+\right.\\&\left.-\frac{f'}{3}+ \frac{f}{6 f'}f'' +
\dot{R} \Theta  \frac{f''^{2}}{ f'} -
  \ddot{R} f^{(3)}- \Theta f^{(3)} \dot{R}- f^{(4)}\dot{R}^2
  \right]\,,\\   
  &\mathcal{G} =-
 \left[ \frac{1}{3}(3 w -1) \mu^{m}+\frac{w}{1+w} \left(f^{(3)}
  \dot{R}^2+  (p^{R}+\mu^{R}) f'+ \frac{7}{3}\dot{R} \Theta f''
+\ddot{R} f'' \right)\right]\,,\\
&\mathcal{H}=-\frac{(w -1) \dot{R} f''}{w +1}\,.
\end{flalign}
\end{subequations}
In the $f(R)=R$ case these equations reduce to the standard equations for the evolution of the scalar perturbations in GR:
\begin{subequations}\label{eqIIordGR}
\begin{flalign}
 &\ddot{\Delta}_{m}^{(k)}-\left(w-{\textstyle\frac{2}{3}}\right) \Theta\dot{\Delta}_{m}^{(k)}
 -  \left[w\frac{k^2}{S^{2}}-\left({\textstyle\frac{1}{2}}+w-{\textstyle\frac{3}{2}}w^2\right)\mu^m\right]\Delta_{m}^{(k)}=0\;,\\
&\mathcal{R}^{(k)}=\left(3w-1\right)\mu^m\Delta_{m}^{(k)}\;.
\end{flalign}
\end{subequations}
If one compares the system \rf{EqPerIIOrd} with the equations for the evolution of scalar perturbations for two interacting fluids in GR one notices that they have the same structure, i.e., one finds friction terms and source terms due to the interaction and  the gravitation of the two effective fluids. It is then natural to ask ourselves if this analogy can be useful to better understand the physics of these models. The answer is affirmative, but with some very important caveats. First of all a more correct way to draw this analogy would be to write the system of equations for  $\Delta_m$ and $\Delta_R=\frac{S^{2}\prjnab^2\mu_R}{\mu_R}$ and analyze their structure rather than using the ones above. Also, as stated in  \cite{SantePert}, one has to be careful in remembering that we are dealing with {\it effective fluids} and, as such, they might violate some basic constraints that standard fluids  usually follow (such as the energy conditions) or present subtleties in the definition of their comoving frame.

However, in spite of  these differences one can still use the coefficients of the $(\Delta_m, \Delta_R)$  equations to  obtain information about the interaction between standard matter and the curvature fluid. In fact, the  coefficients of these equations are found to behave as a ratio of polynomials in the wavenumber and have a non trivial behavior in $t$.  This kind of behavior is very different to what is found in a GR-two fluid system. In a photon-baryon system, for example, the dissipation terms grow as $k^2$ and  behave like $1/t$ in time. Therefore, unlike the Thompson scattering in the baryon-photon system, the effect of the interaction between matter and non-linear gravitation can influence large and small scales alike, depending on the structure of the action. Hence, from the distribution of the structures in the observed sky one can deduce constraints on the nature of the theory of gravity. Another important difference with GR is that  \rf{EqPerIIOrd}  are  scale dependent for any value of the barotropic factor. This means that whatever the equation of state of standard matter, the perturbation solutions will always depend on the scale at which they are calculated. This is true even in the special case of dust which in GR is associated to a scale invariant perturbation.

An additional interesting feature is the meaning of the long and short wavelength limit in \rf{EqPerIIOrd}. As it is clear from \rf{eqIIordGR}, in GR these  limits  can be defined  by comparing  the values of the quantity  $\frac{k^2}{S^2 \Theta^2}$ and the ratio between matter term in the $\Delta_{m}$ coefficient and $\Theta^2$. However, looking at \rf{EqPerIIOrd} it is clear that  the situation here is more delicate. For example, the short wavelengths regime cannot be defined  as simply $\frac{k^2}{S^2 \Theta^2}\gg 1$, but  $\frac{k^2}{S^2 \Theta^2}$  has to be bigger than {\it all} the other quantities appearing in the coefficients $\mathcal{B}$, $\mathcal{C}$ and $\mathcal{F}$ divided by $\Theta^2$. The same reasoning holds for  the long wavelengths: $\frac{k^2}{S^2 \Theta^2}$  has to be smaller than {\it all} the other quantities appearing in the coefficients $\mathcal{B}$, $\mathcal{C}$ and $\mathcal{F}$ \footnote{However, there is a difference between the two limits because in the long wavelength limit one can always set $\left|\frac{k^2}{S^2 \Theta^2}\right|\approx0$, while in the short wavelength the definition is completely dependent on the values of the coefficients and, as consequences, on the features of the background.}.  In this sense the structure of \rf{EqPerIIOrd} suggests the presence of  a least three different regimes in the evolution of the perturbations. Firstly,  the ``deep super-horizon" regime in which $k$ is effectively zero, an intermediate one (or two depending on the value of the barotropic factor $w$) which is determined by the details of the background and a ``deep subhorizon" regime in which one has effectively $k\rightarrow\infty$. 
The situation seems to become even more complicated when the fourth order gravity action posses dimensional constants (like in the case $f(R )=R +\alpha R^{n}$). Since these constants are associated with the scales at which the different contributions to the action become dominant, one would expect  the introduction of additional scales into the theory, i.e., further possible evolution regimes for scalar perturbations.  
\section{Examples}
The considerations drawn above are the most that we can extract from the  system of perturbation equations in its general form \rf{EqPerIIOrd}. In order to  obtain more information we need to apply these equations to some specific models. The main issue related to this, is the lack of exact or numerical backgrounds in the most popular models of $f(R)$-gravity. This is due on a side to the complexity of the equations which makes very difficult their resolution, on the other to the difficulty in findings constraints on the values of the parameters which are needed to perform numerical integration.  The  error due to the use of a background which is not a solution of the cosmological equations is very difficult to estimate and it could well depend on the detailed of the model considered. For this reason we will investigate here some very simple models for which some backgrounds are known in order to gather as much information as possible and test our general conclusions.
\subsection{The case of  $R^{n}$-gravity.}
Let us consider the case  $f(R)=\chi R^{n}$, called also sometimes $R^{n}$-gravity, whose action reads
\begin{equation}\label{lagrRn}
L=\sqrt{-g}\left[\chi R^{n}+2{\cal L}_{m}\right]\;,
\end{equation}
where ${\cal L}_{m}$ is the matter Lagrangian. The above theory constitutes the simplest possible example of fourth-order gravity. In this toy model  the cosmological equations associated with  a  Friedmann-Lema\^{\i}tre-Robertson -Walker (FRLW) metric are particularly easy to deal with. 
The FLRW dynamics of this model has been investigated via a complete
phase space analysis in \cite{Carloni:2004kp}. This analysis shows
that for specific intervals of the parameter $n$ there is a set of
initial conditions with non-zero measure for which the cosmic
histories include a transient decelerated phase which evolves
towards an accelerated expansion one. This first phase, characterized by 
\begin{equation}\label{bckgRnFRW}
 S=S_0 t^{\frac{2n}{3(1+w)}}\;,\qquad K=0\;,\qquad \mu^m=\mu^m_{0} t^{-2n}\;,
\end{equation}
was argued to be suitable for the structure formation to take place. For this reason we will use \rf{bckgRnFRW} as background.

\subsubsection{The long wavelength limit}
Let us start analyzing the evolution of the long wavelength perturbations during the phase \rf{bckgRnFRW}.
In the long wavelength limit the wavenumber $k$ is considered to be so small
that the wavelength $\lambda=2\pi S/k$ associated with it is much
larger than the Hubble radius. Equation \rf{CovHarmDef} then
implies that all the Laplacians can be neglected and the spatial
dependence of the perturbation variables can be factored out.  
In GR, one can prove \cite{Conserved} that  in this limit and in spatially
flat ($K=0$) backgrounds the equations for the variable C reduces to $\dot{C}=0$
i.e. the variable $C$ is conserved. From the structure of equations \rf{eqC} above
 one can see that in this limit and in spatially
flat ($K=0$) backgrounds the \rf{eqCHarm} reduces to $\dot{C}=0$
i.e. the variable $C$ is conserved also in the $f(R)$ case. This means that the number 
of equations  in \rf{SysIOrdHarm} can be reduced to three.

\begin{figure}[htbp]
\includegraphics[scale=1.2]{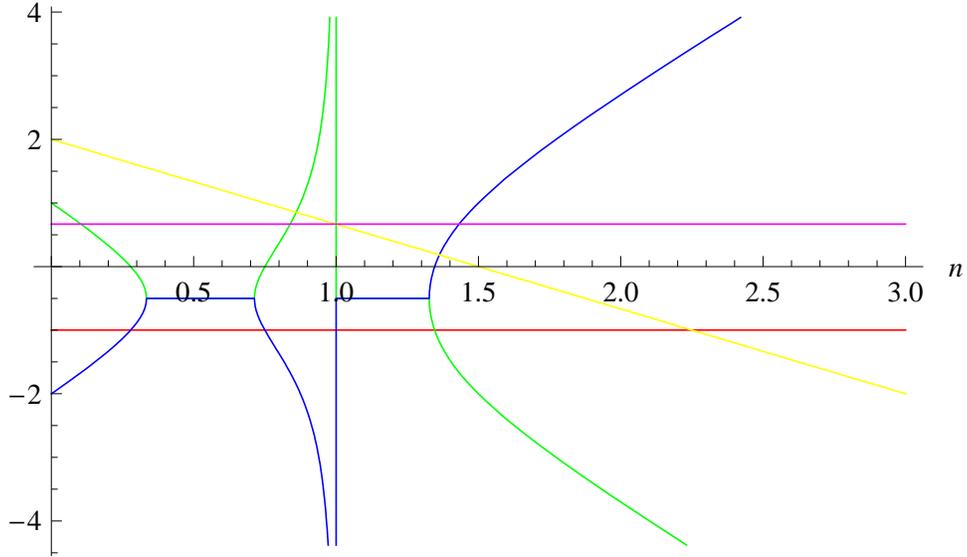}
\caption{Plot against $n$ of the real part of the long wavelength modes
for $R^n$-gravity in the dust case (blue, red green and yellow lines) together
with the GR modes (red and purple line) \cite{SantePert}.}
 \label{plotdust}
\end{figure}

Substituting the background \rf{bckgRnFRW} in the equations thus obtained and decupling 
the density perturbation equation one obtains:
\begin{eqnarray}\label{EqDelta3Ord}
(n-1)\dddot{\Delta}_m-(n-1)\left(\frac{4 n w }{w
+1}-5\right)\frac{
   \ddot{\Delta}_m
   }{t}+ \mathcal{D}_1 (n,w)\;\frac{\dot{\Delta}_m}{t^2}+\mathcal{D}_2 (n,w)\frac{\Delta _m}{ t^3}+\mathcal{D}_3(n,w)\;\mathcal{C}_0\; t^{-\frac{4 n}{3 (w +1)}-1}=0\;,
\end{eqnarray}
where, $C_0$ is the conserved value for the quantity $C$ and
\begin{eqnarray}
  \mathcal{D}_1  (n) &=& -\frac{2 \left(-9 (2 (n-1) n+1) w ^2+6 n (n (4 n-7)+1)
   w +18 w +n (4 n (8 n-19)+33)+9\right) }{9 (w
   +1)^2}\;, \\
\mathcal{D}_2(n) &=&\frac{2 ((2 n-1) w -1) (4 n-3 (w +1))
(3 (w +1)+n
   (-9 w +n (6 w +8)-13)) }{9  (w +1)^3
  }\;,\\
\mathcal{D}_3(n) &=& -\frac{n (21 w -6 n (w +2)+31)-18
(w +1)
   }{6  S_0^2}\;,
\end{eqnarray}
 This equation admits the general solution
\begin{eqnarray}\label{soluzione delta gen LWL}
   &&\Delta_m=K_1 t^{\frac{2 n w
   }{w +1}-1}+K_2 t^{\alpha _+}+K_3 t^{\alpha _-}-K_4 \frac{\mathcal{C}_0}{S_0^2}
   t^{2-\frac{4 n}{3 (w+1)}}\;,
\end{eqnarray}
where
\begin{eqnarray}
&&\alpha_{\pm} = -\frac{1}{2}+\frac{n w }{w +1}\pm \frac{\sqrt{(n-1)  \left(4 (3
   w +8)^2 n^3-4 (3 w  (18 w +55)+152) n^2+3 (w +1) (87
   w +139) n-81 (w +1)\right)}}{6 (n-1) (w
   +1)^2}\;,\nonumber\\\\
 &&K_4= \frac{9 (w +1)^3 (18 (w +1)+n (-21 w +6 n (w +2)-31))}{8 (n
   (6 w +4)-9 (w +1)) \left(6 (w +2) n^3-(9 w +19) n^2-3
   (w +1) (3 w +1) n+9 (w +1)^2\right)}\;.
\end{eqnarray}
Let us now focus on the case of dust ($w=0$). The above solution
becomes
\begin{eqnarray}\label{soluzione delta gen LWL dust}
   &&\Delta_m=K_1 t^{-1}+K_2 t^{\alpha _+|_{w=0}}+K_3 t^{\alpha _-|_{w=0}}-K_4 \frac{\mathcal{C}_0}{S_0^2}
   t^{2-\frac{4 n}{3}}\;,
\end{eqnarray}
where
\begin{eqnarray}
 &&\alpha_{\pm}|_{w=0} = -\frac{1}{2}\pm\frac{\sqrt{(n-1) (n (32 n (8 n-19)+417)-81)}}{6 (n-1)}\;,\\
 &&K_4|_{w=0}= \frac{9 (n (12 n-31)+18)}{8 (4 n-9) \left(12 n^3-19 n^2-3 n+9\right)}\;.
\end{eqnarray}
A graphical representation of the behavior of the exponent of the
modes of \rf{soluzione delta gen LWL dust} as $n$ changes is given
in Figure \ref{plotdust}. This solution has many interesting
features. For $0.33 < n < 0.71$ and $1 < n < 1.32$ \footnote{The
values of $n$ presented here and in the following are the result of
the resolution of algebraic equations of order greater than two.
These values have been calculated numerically. This
means that the values we will give for the interval $n$ will be
necessarily an approximation.} the modes $t^{\alpha_{\pm}|_{w=0}}$
become oscillatory. However, since the real part of the exponents
$\alpha_{\pm}|_{w=0}$ is always negative, the oscillation are damped
and bound to become subdominant at late times. The appearance
of this kind of modes is not associated with any peculiar behavior
of the thermodynamic quantities in the background i.e. none of the
energy condition are violated for the values of $n$ which are associated with the oscillations. The nature of these oscillations is then an higher
order phenomenon.  Also, for most of the values of $n$ the
perturbations grow faster in $R^n$-gravity than in GR. In fact only
for $1.32\leq n < 1.43$ all the modes grow with a rate slower than
$t^{2/3}$.

Probably the most striking feature of the solutions \rf{soluzione
delta gen LWL dust} and  \rf{soluzione delta gen LWL} is that the
long wavelength perturbations can grow if
the Universe is in a state of accelerated expansion (see Figure
\ref{plotdust}). The consequence
of this feature is quite impressive because it implies, against our 
(GR-based) intuition, that in $R^{n}$ gravity large scale structures can 
in principle also be formed in accelerating backgrounds. The suppression of
perturbations due to the presence of classical forms of Dark Energy
(DE) is one of the most important sources of constraints on the
nature of DE itself. Our example shows that if one considers DE as a
manifestation of the non-Einsteinian nature of the gravitational
interaction on large scales, there is the possibility to have an
accelerated expanding background  that is compatible with the growth
of structures. 

In the limit $n\rightarrow 1$ two of the modes of \rf{soluzione
delta gen LWL} reproduce the two classical modes $t^{2/3}$ and
$t^{-1}$ typical of GR, but the other two diverge. At first glance
this might be surprising,  but it does not represent a real pathology
of the model. In fact, when $n=1$, equation \rf{EqDelta3Ord} reduces to a first
order differential equation with a forcing term, which presents a two modes solution. This implies that in this case the additional modes in the solution \rf {soluzione delta gen LWL}  can be discarded and GR is recovered.

Solving for the other variables we can also obtain the solution
for the other scalars:
\begin{eqnarray}\label{soluzione others gen LWL}
   && \mathcal{R}=K_5 t^{\frac{2 n w
   }{w +1}-3}+K_6 t^{\beta _+}+K_7 t^{\beta _-}-K_8 \frac{\mathcal{C}_0}{S_0^2} t^{-\frac{4 n}{3 (w
   +1)}}\\&&\Re= K_9 t^{\frac{2 n w
   }{w +1}-1}+K_{10} t^{\gamma _+}+K_{11} t^{\gamma _-}-K_{12} \frac{\mathcal{C}_0}{S_0^2}  t^{-\frac{4 n}{3 (w
   +1)}-1}
\end{eqnarray}
where
\begin{eqnarray}
  \beta_{\pm} &=& \alpha_{\pm}-2\;,\\
 \gamma_{\pm} &=& \alpha_{\pm}-3\;,
\end{eqnarray}
and the constants $K_5,..K_{12}$ are all functions of $K_1,..K_4$.
These expressions are rather complicated and will not be given here. It is
interesting that these quantities have an oscillatory behavior for the
same values of $n$ for which $\Delta_m$ is oscillating. Also for
these quantities the oscillating modes are always decreasing.

\subsubsection{Short wavelengths.}
The analysis of the properties of the small scale perturbations is more complicated than long wavelength ones. This because, if, as often happens, it is not possible to obtain exact solutions in this regime one is bound to find other means the relevant physical information. 

A way of doing that, based on a statistical treatment of the  behavior of the perturbation at different scales, is to define some characteristic quantities which help extract physical content
from these solutions. One of the most important of these quantities is the power spectrum $P_X(k)$ of a perturbation variable $X$. Mathematically $P_X(k)$ is defined as the variance of the
amplitudes of the Fourier transform of $X$ at a given value of $k$ \cite{Coles}:
\begin{equation}
\langle X({\mathbf k_1})X({\mathbf k_2})\rangle=P_X(k_1) \delta({\mathbf k_1}+{\mathbf k_2})\;,
\end{equation}
where ${\mathbf k_i}$ are two wavevectors characterizing two Fourier components of $X$ and $P({\mathbf k_1})=P(k_1)$ because of isotropy in the distribution of the perturbations. The power spectrum $P_X(k)$ tells us how the fluctuations associated to $X$ depend on the wavenumber at a specific time and carries information about the amplitude of these perturbations (but not on their spatial structure). Its features are directly connected with the structure of the equations. For example, \rf{eqIIordGR} shows that in the case of GR and pure dust the matter fluctuations equation does not depend on $k$ , so the matter power spectrum can be considered constant (i.e. scale invariant) \cite{PaddyPert}. 

A spectrum of one of the perturbation variables is a powerful tool for comparing the
predictions of the system \rf{EqPerIIOrd} with observations and is able
to reveal a great deal of information on the physics of scalar perturbations. In what follows we will focus on the matter perturbation spectrum $P_\Delta(k)$ only, because this spectrum is the one that can be directly measured via large scale surveys. Our spectra will be normalized in such a way that it is unity on super-horizon scales. This normalization can then be scaled with current observations of the power 
spectrum on large scales (see \cite{WMAP} for the latest constraints). We will also analyze the time variation of this spectrum in order to obtain information on the ways in which the perturbations evolve in time on different scales. 

\begin{figure}[htbp]\label{PkRn}
\subfigure[Plot of  $P_\Delta (k)$ at $\tau=1$ for $R^n$-gravity and $n>1$. Note that the spectrum is composed of three parts corresponding to three different evolution regimes for the perturbations. ]{\includegraphics[scale=0.75]{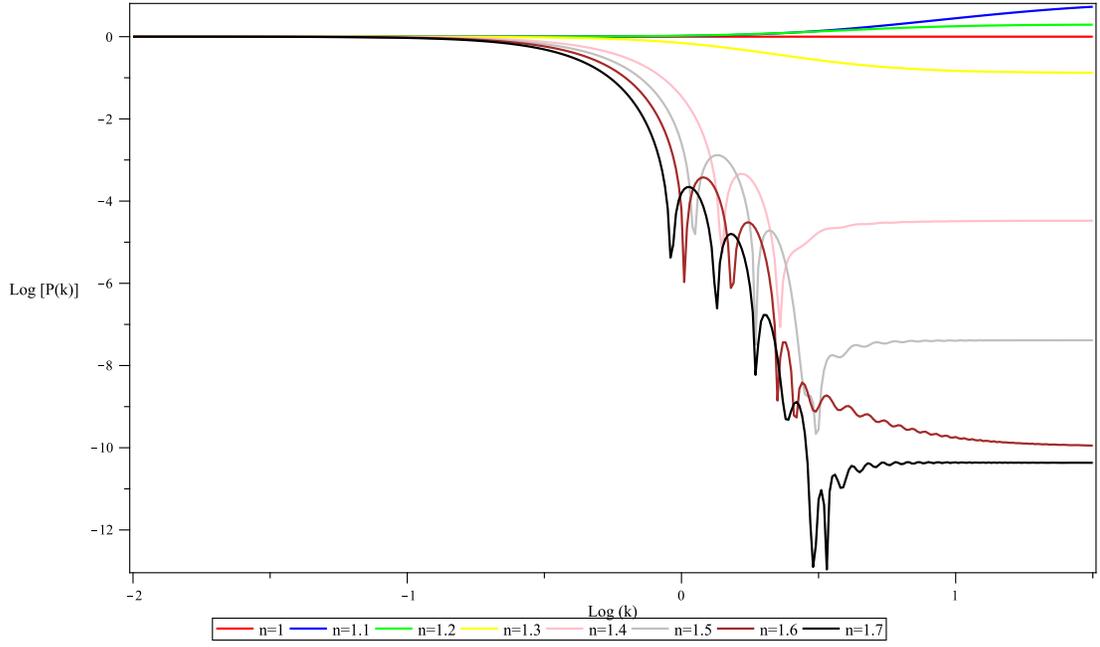}\label{Fig11}}
\subfigure[Plot of the $P_\Delta (k)$ as a function of $k$ for $R^n$-gravity at $\tau=1$ and $0<n<1$. The spectra for $n=0.6$ and $n=0.5$
only approach the scale invariant plateau at extremely high $k$ when compared to the other curves. Note the behavior of the spectrum for $n\approx 0.8$, differently from all the other cases, there is basically no loss of power in the spectrum at large $k$. ]{\includegraphics[scale=0.50
]{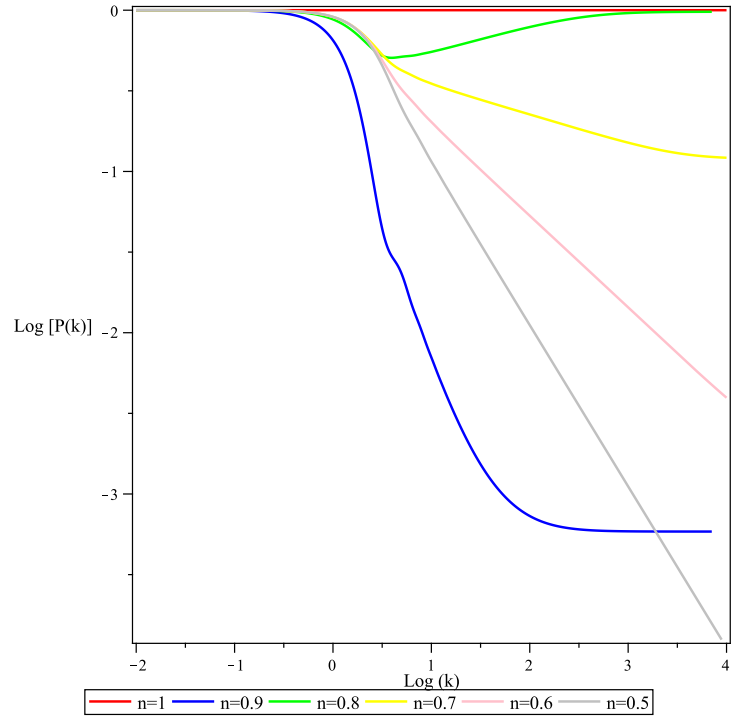}\label{Fig12}}
\caption{Matter power spectra for $R^n$-gravity (from \cite{StructForm})}
\end{figure}

\begin{figure}[htbp]
\begin{center}
\includegraphics[scale=0.50]{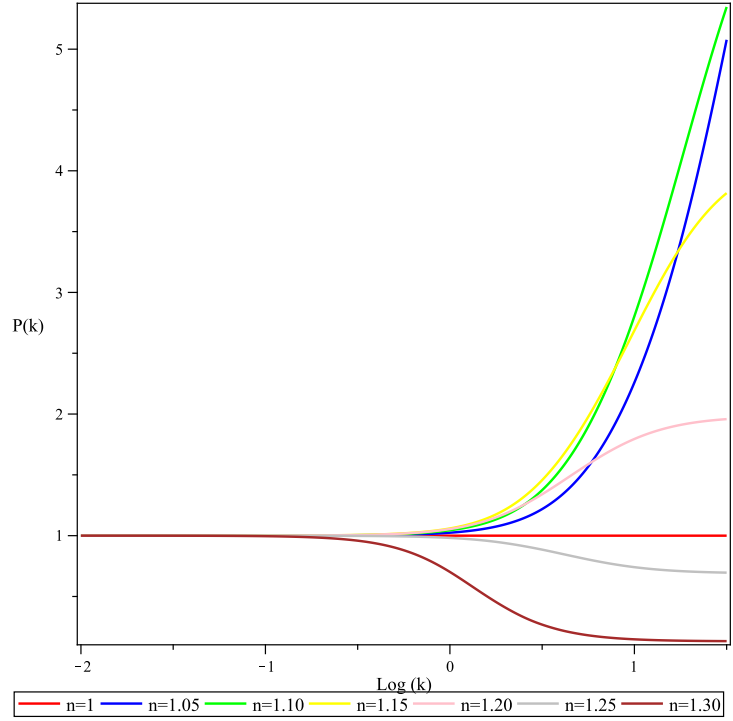}
\caption{Detail of the plot of  $P_\Delta (k)$ for $R^n$-gravity at $\tau=1$. As expected, for these values of $n$ we find the presence of the three regimes mentioned in the text. Note also that for $n\approx 1^{+}$ the small scales are characterized by an excess of power \cite{StructForm}. Such features are compatible with some of the results found in  \cite{otherperts}. \label{Fig13}}
\end{center}
\end{figure}

Let us then consider $P_\Delta(k)$ for \rf{EqPerIIOrd} at $\tau=S/S_0=1$
(Figures \ref{Fig11}, \ref{Fig12}, \ref{Fig13}).  As mentioned above  the $k$-structure of Equations \rf{EqPerIIOrd}  suggests that in fourth order gravity there exist at least three different growth  regimes  of the perturbations. This is confirmed by our results for this example. In particular, in the case of
dust and for any values of the remaining parameters we have that:  (i) on very large scales the spectrum
goes like GR i.e. it is scale invariant;  (ii) as $k$ becomes bigger the scale invariance is broken and oscillations in the spectrum appear;  (iii)   for even larger $k$ the spectrum becomes again scale invariant. However, on these scales the spectrum can contain either an excess or deficit of power depending on the value of  $n$. In particular for $n\approx 1^+$ small scales have more power than large scales,
but, as one moves towards larger values of $n$, the small scale modes are suppressed.
For $0<n<1$, instead the drop in power seems to decrease as one moves from $n=0$ towards
$1^-$ and we see a sudden increase for $n\approx 1^-$. It is worth noting
the case $n\approx 0.8$ for which there is basically no difference in power between
large and small scales and there are no significant oscillations in the spectrum.

\begin{figure}[htbp]
\begin{center}
\includegraphics[scale=0.80]{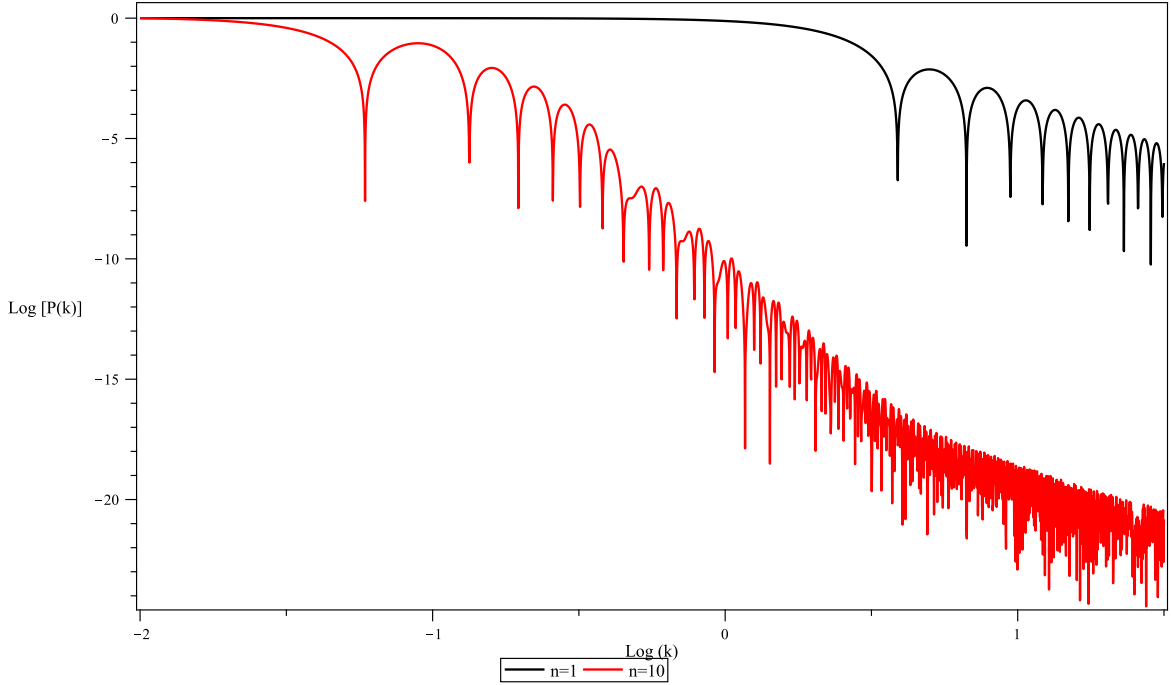}
\caption{$P_\Delta (k)$ for $R^n$-gravity at $\tau=1$ in the
case of radiation for GR and $R^n$-gravity with $n=10$. As expected in this last plot  we find four different
regimes instead of the three of the dust case: a first regime for $k\rightarrow0$ which is scale invariant; a second and a third regime which correspond to the two different slopes between $k\approx 10^{-2}$ and  $k\approx 10^{1/2}$ and a fourth regime which has the same slope of the GR plot  \cite{StructForm}. \label{Fig14}}
\end{center}
\end{figure}

Further  indication of the link between the $k$ structure of \rf{EqPerIIOrd} and the different regimes of the matter power spectrum can be seen if one analyzes  the power spectrum in a radiation dominated era. In this case the perturbations equations contain an additional $k$-term which is not present in the dust case.  This means that one would expect {\it four} different regimes, rather than three. In Figure \ref{Fig14} we have plotted the matter power spectrum of $R^n$-gravity in the case of radiation and $n=10$ (this value of $n$ is chosen only for convenience), and as expected one can recognize four different regimes. Note that we obtain the same $k$ scaling as in GR on small scales.

\begin{figure}[htbp]
\begin{center}
\includegraphics[scale=0.5]
{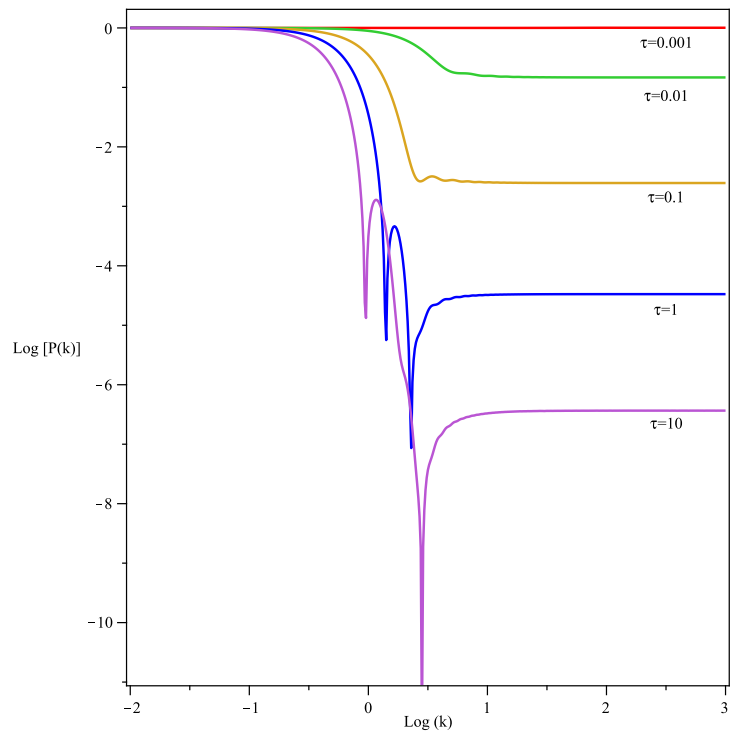}
\caption{Evolution of $P_\Delta (k)$ for $R^n$-gravity for $n=1.4$. The spectrum has been normalized in such a way  that the curves coincide at large scales. Note how, as time passes, small scale perturbations are dissipated and oscillation appear  \cite{StructForm}. \label{Fig15}}
\end{center}
\end{figure}

Finally, further information on the dynamics of the matter perturbations can be obtained examining the time evolution of the power spectrum. In Figure  \ref{Fig15} we give the power spectrum for $n=1.4$ at different times. One can see that, as the universe expands, the small scale part of the spectrum is more and more suppressed and oscillations start to form.  On the other hand the large scales do not seem to be evolving, which might appear in contrast with what mentioned above. However,  this is a byproduct of the normalization:  for clarity we have normalized the spectrum in such a way that every curve has the same power in long wavelength limit.

The features of the spectra that we have derived can be then interpreted in terms of the interaction between the curvature fluid and standard matter \cite{StructForm}. On very large and very small scales, the coefficients of equations \rf{EqPerIIOrd}   become independent from $k$ so that the evolution of the perturbations does not change with the scale and the power spectrum is scale invariant. On intermediate scales the interaction between the two fluids  is maximized and the curvature fluid acts as a relativistic component whose pressure is responsible for the oscillations and the dissipation of the small scale perturbations in the same way in which the photons operate in a baryon-photon system\footnote{This suggests the following interesting interpretation for the perturbation variables $\mathcal{R}$ and $\Re$. These quantities can be thought to represent  the modes associated with the contribution of the additional scalar degree of freedom typical of $f(R)$-gravity. In this sense the spectrum can be explained physically as a consequence of the interaction between  these scalar modes, which function as a scalar gravitational wave, and standard matter.}. The result is a considerable loss of power for a relatively small variation of the parameter $n$. For example, in the case $n=1.4$ the difference in power between the two scale invariant parts of the spectrum for $n=1.1$ is of  one order of magnitude while for $n=1.6$  is about ten orders of magnitude.

Probably the most important consequence of the form of the spectrum presented above is the fact that the effect of these type of fourth order corrections is evident only for a special range of scales, while the rest of the spectrum has the same $k$ dependence of GR (but different amplitude). This implies that we have a spectrum that both satisfies the requirement for scale invariance and has distinct features that one could in principle detect, by combining future Cosmic  Microwave Background (CMB) and large scale surveys (LSS) \cite{Planck, SDSS}.

\subsection{The case $f(R)=R+\alpha R^{n}$}\label{SecRRn}

Let us now consider an $f(R)$ theory characterized by the following action,
\begin{equation}\label{lagr RRn}
L=\sqrt{-g}\left[R+\alpha R^{n}+{\cal L}_{M}\right]\;.
\end{equation}
This theory has gained much popularity as a fourth order gravity model within the
context of both inflation and dark energy \cite{Ottewill, star80, Suen, Carroll}.

Unlike $R^{n}$-gravity, \rf{lagr RRn} includes explicitly the
Hilbert-Einstein term, so that one can consider it as the result of an additive correction to GR.
As a consequence this model  posses an additional physical scale which is related, in our units, to the coupling constant $\alpha$ by the relation $L=\alpha^{2(n-1)}$, making \rf{lagr RRn} the simplest fourth order
gravity theory with an additional scale for the gravitational interaction.

In the following we will take $\alpha$, which in our units is the ratio between the coupling constant of
the fourth order corrections, to be positive definite. Of course, we expect this model
to behave much in the same way as the model discussed in the previous section for $\alpha\rightarrow\infty$,
and to recover GR for $\alpha=0$. This also means that any new feature in this model will emerge for intermediate values of the coupling.

The $f(R)=R+\alpha R^{n}$ model has been analyzed at the level of the background using
many different approaches (see for example \cite{Ottewill, star80, Suen, Carroll, Teyssandier:1989dw}),
but probably some of the most interesting results for cosmology have been found using the dynamical system
approach \cite{Barrow Hervik,SanteGenDynSys}. The dynamical systems analysis proved
that this class of models has, like $R^{n}$-gravity, an unstable fixed point associated
with the Friedmann-like solution\footnote{Unlike the previous example, the structure of the phase space for this model is not well known. This means that, although in this theory one has a fixed point that resembles the one of $R^n$-gravity, it is not obvious that it plays the same role. In addition, this background is not, in general, a physical solution of the cosmological equations \cite{SanteGenDynSys}. This means that there are cosmic histories in which the general integral of these equations approximates the behavior we consider, but it will never be exactly the same. We choose to treat this as a further approximation in our investigation. }  $a=t^{2n/3(1+w)}$.

\begin{figure}[htbp]
\subfigure[ Plot of $P_\Delta(k)$ for $n>1$, $\alpha=10$ and various values of $k$]{\includegraphics[scale=0.45]{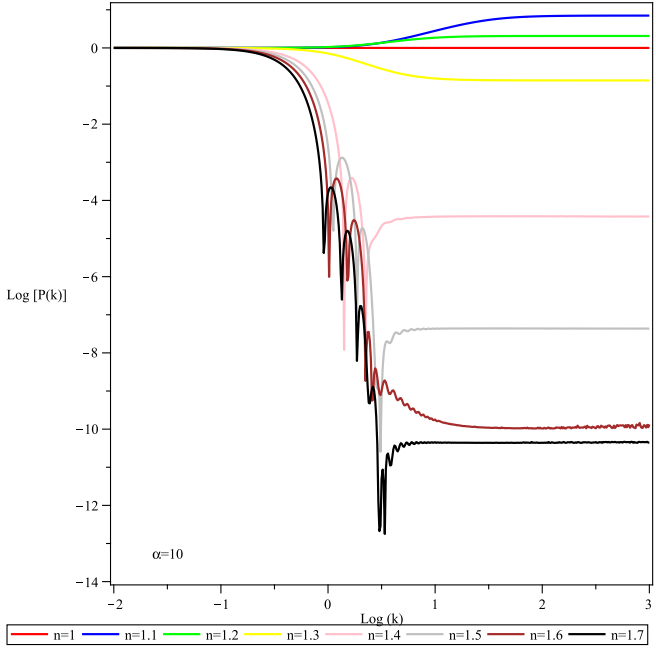}\label{PkRRnX1}}
\subfigure[ Plot of $P_\Delta(k)$ for $n>1$, $\alpha=1$ and various values of $k$]{\includegraphics[scale=0.45]{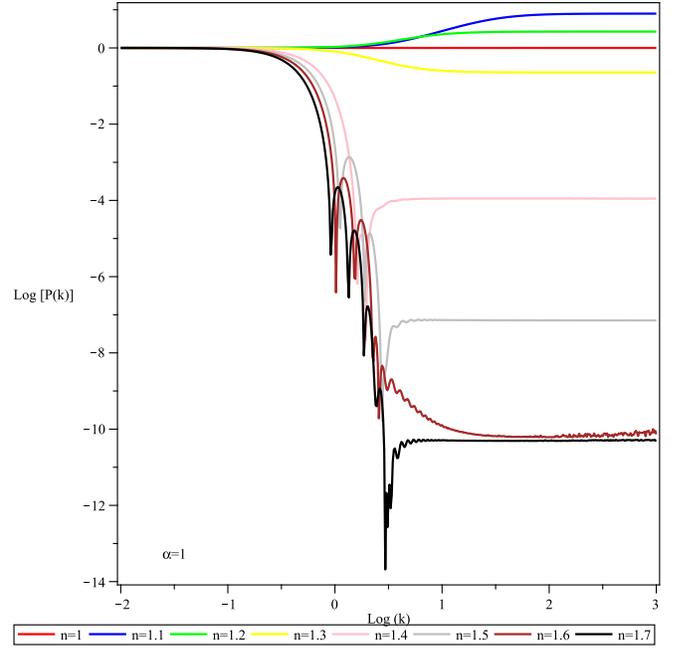}\label{PkRRnX2}}
\subfigure[ Plot of $P_\Delta(k)$ for $n>1$, $\alpha=0.1$ and various values of $k$]{\includegraphics[scale=0.45]{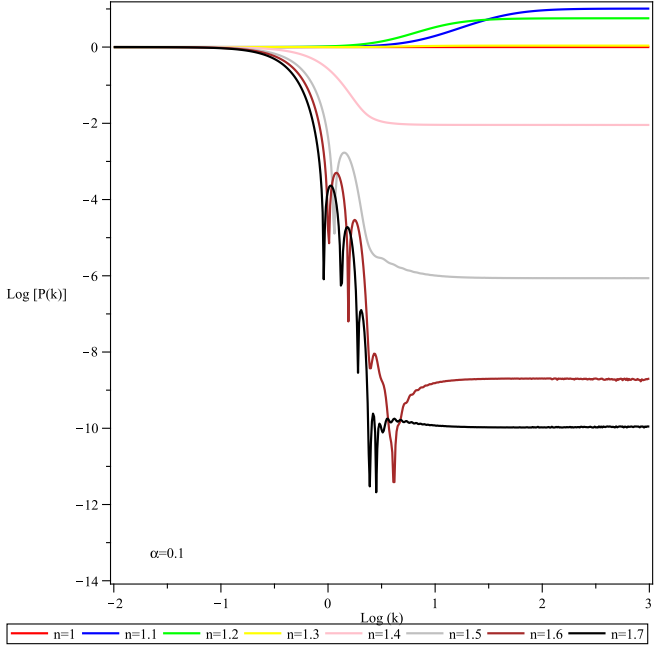}\label{PkRRnX3}}
\subfigure[ Plot of $P_\Delta(k)$ for $n>1$, $\alpha=0.01$ and various values of $k$]{\includegraphics[scale=0.45]{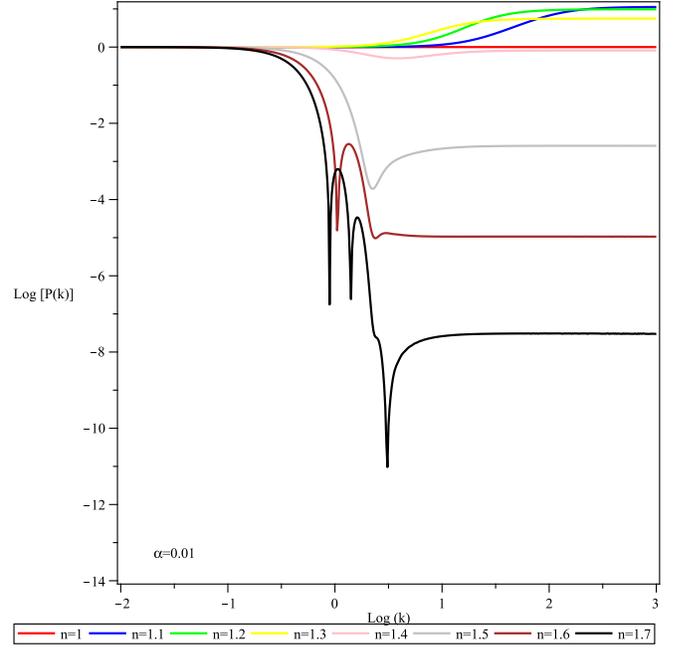}\label{PkRRnX7}}
\caption{Plot of  $P_\Delta (k)$ as a function of $k$ for $R+\alpha R^n$-gravity at $\tau=1$ for $n>1$  \cite{StructForm}.}\label{PkRRn}
\end{figure}

One can then analyze the behavior of the perturbations of this background in $f(R)=R+\alpha R^{n}$ and compare it with the one found in $R^n$-gravity. However, dealing with this model  is considerably more
complicated than dealing with $R^{n}$-gravity. For example, one is unable to find exact
solutions for the perturbation equations, even in the long wavelength
limit. For this reason in the following we will focus directly on the features of the power spectrum  which are shown for various values of the parameters $\alpha$ and $n$ in Figures \ref{PkRRn}.  As one can clearly see, these plots resemble the ones we have  derived in the previous example. There are, of course, differences in the position of the oscillations and the amount of the power drop, but one finds again three different regimes in the case of dust and two of them ($k\rightarrow 0$ and $k\rightarrow \infty$ ) correspond to scale invariance. Particularly interesting is the fact that in principle the values of $\alpha$ and $n$ can be fine tuned in such a  way  to obtain a spectrum in which the small scales have the same power as the large ones. In a situation like this most of the spectrum would be scale invariant and all the deviations would be concentrated around a specific scale.

\begin{figure}[htbp]
\subfigure[Plot of $P_\Delta(k)$ for $n=1.4$, $\alpha=10$ and evaluated at various values of $\tau$]{\includegraphics[scale=0.40]{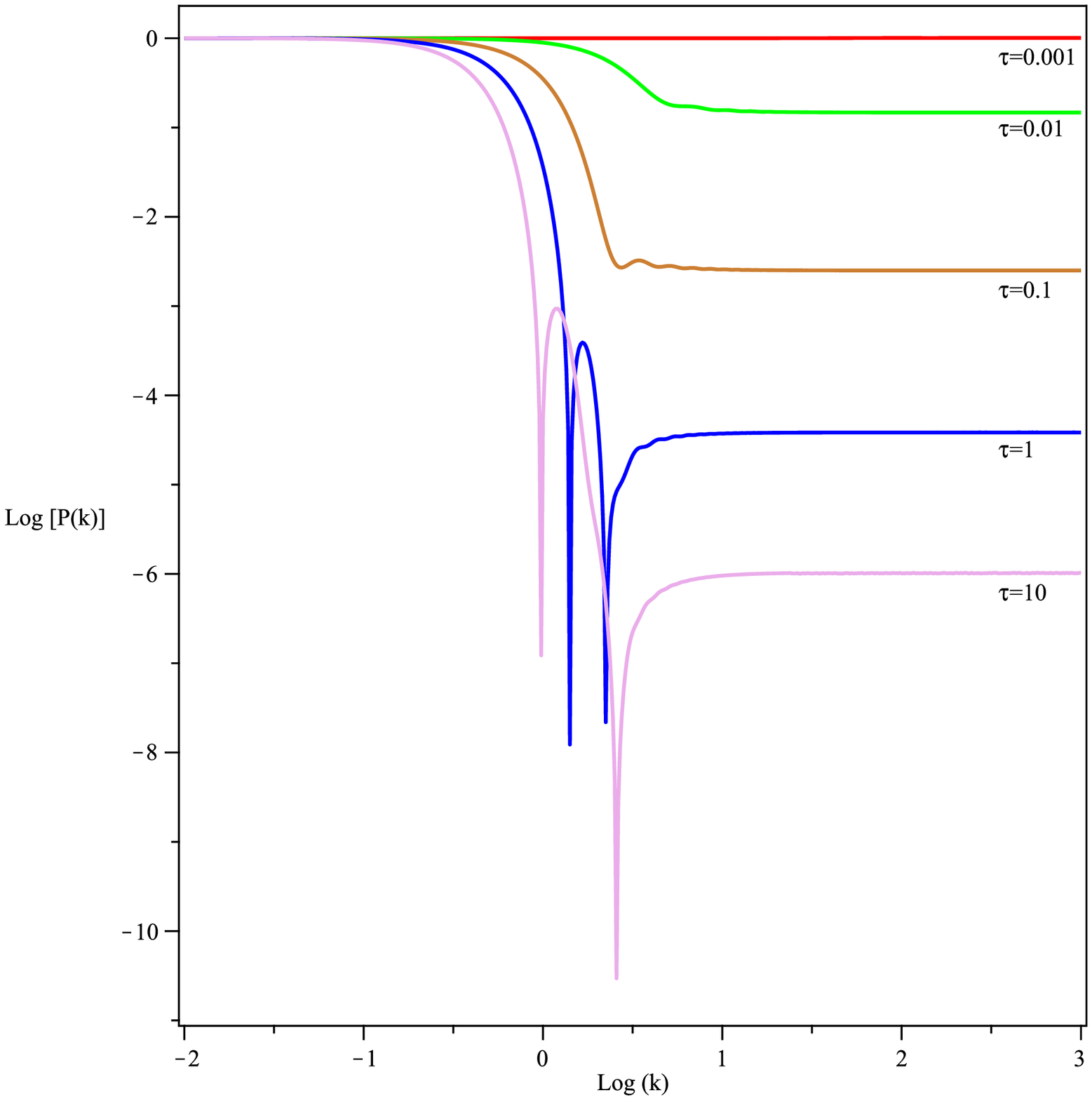}\label{Fig33}}
\subfigure[Plot of $P_\Delta(k)$ for $n=1.4$, $\alpha=0.01$ and evaluated at various values of $\tau$]{\includegraphics[scale=0.40]{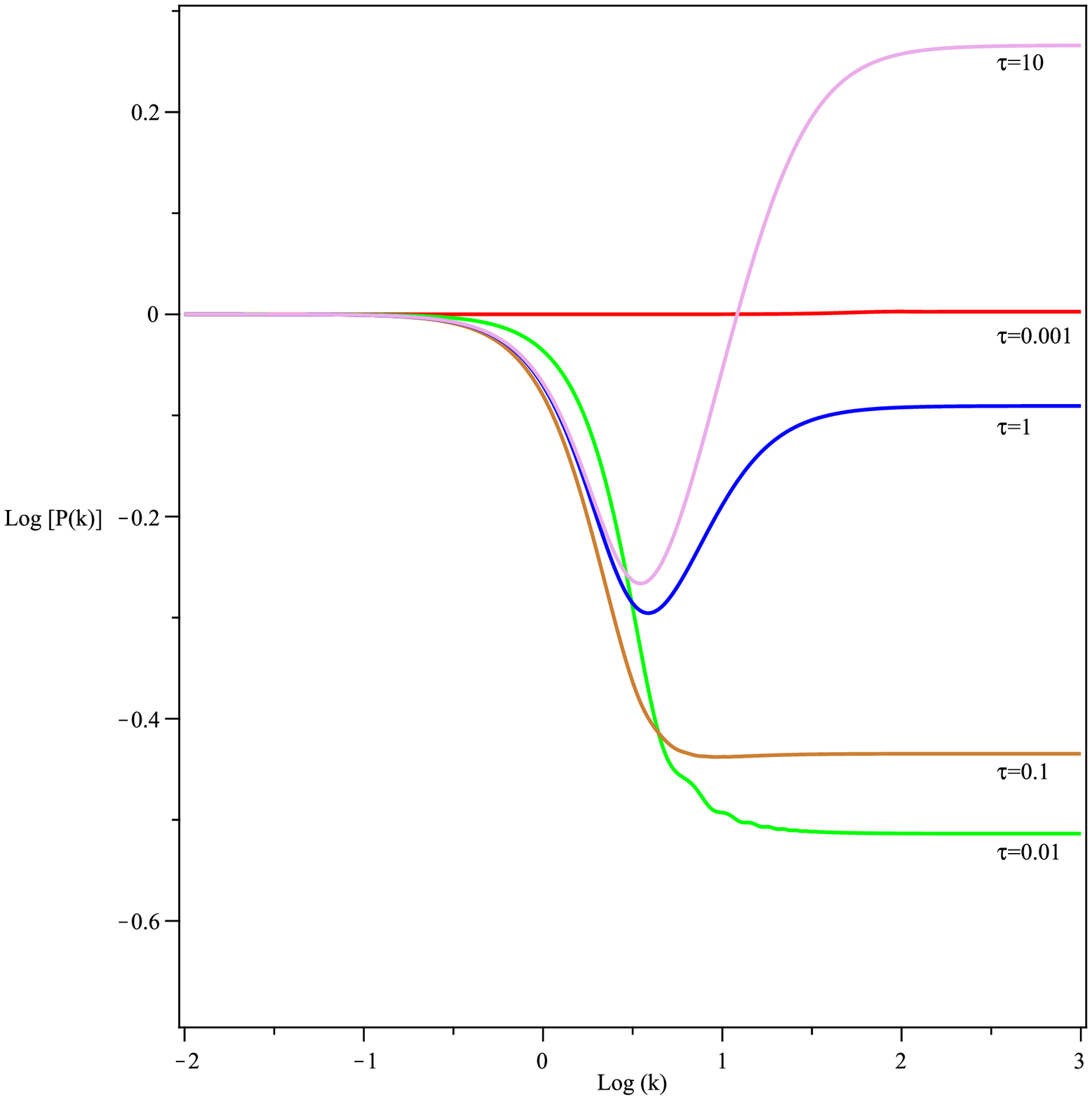}\label{Fig34}}
\caption{The time evolution of the matter power spectrum in $R+\alpha R^n$-gravity for $n=1.4$ and differing values of $\alpha$. Note the
drastically differing vertical scales in the plots. Additionally, note the rise in power at small scales at late times in the case of
$n=1.4$ and $\alpha=0.01$  \cite{StructForm}.}
\end{figure}

The time evolution of these spectra also reveals some interesting insights into the dynamics of the matter fluctuations. As usual for large values of $\alpha$ the evolution is very similar to the one obtained for $R^n$-gravity  as it is shown in Figure \ref{Fig33}. However, when the value of the coupling changes the behavior of the perturbations can change dramatically. An example is given in figure \ref{Fig34} representing the power spectrum of the model $(n=1.4,\alpha=0.01)$ in which the small scale perturbations are first dissipated and successively start to grow again. This means that in principle one could choose $n$ and $\alpha$, such that for example the small scale perturbation grow at different rates at different times. This property could be useful in  the resolution of open problems in GR structure formation, like the cosmological dark matter or the excess of dwarf red galaxies.

Therefore, in spite of the presence of an additional scale, the power spectrum in this class of model seems to preserve most of the main structure of the one in $R^n$-gravity. This implies  that all the considerations made in the previous section concerning the physical mechanisms behind the form of the spectrum  can be made also in this case. At this point the natural question that arises is about the generality of this result. Is it possible that we have found a characteristic feature of $f(R)$-gravity? Unfortunately, due to the lack of exact background solutions (many of the more complicated models of $f(R)$ do not admit power law solutions) and the difficulties related with the numerical resolution of the perturbation equations, one is unable to answer directly to this question. However, the important point is that if, as we have argued in the previous sections, the structure of the power spectrum is determined only by the $k$-structure of the  perturbation equations, one can make the conjecture that the features we have observed in our two simple models are indeed common to all the $f(R)$ models. If this is confirmed then we would have found a crucial method to test the existence of higher order corrections and it would be a powerful constrain for the parameters of any $f(R)$ model. Another issue concerns the relevance of the form of the background in emergence of the features listed above. Can specific choices of background influence the presence of this characteristic feature? In order to answer this question we will consider another simple example: a de Sitter Universe. 

\subsection{Perturbations of the de Sitter spacetime in a general $f(R)$-gravity.}
Let us consider now the cosmological perturbations around a de Sitter background in $f(R)$-gravity.  The presence of a de Sitter background in $f(R)$-gravity is one of the most important features of these theories because it has the potential to model both inflation and dark energy \cite{Nojiri:2007as,Cognola:2007zu}. In fact,  it has been proven that a viable $f(R)$-gravity model unifying inflation  and late time acceleration in the form of double de Sitter solution can be always constructed numerically \cite{Cognola:2008zp}.  As we will see, however, such backgrounds are not suitable for structure formation, because matter is dissipated very quickly. Notwithstanding this physical issue, the peculiar properties of this metric allow us to  go deeper in the understanding of the perturbation equations.

Let us consider a Universe in which the background is given by a de Sitter spacetime characterized by a scale factor $S=S_0 e^{\beta t}$ and vacuum ($\mu^m=0$). Substituting in the cosmological equations it is easy to show that $\beta$ has to satisfy the equation
\begin{equation}\label{dSCond}
18 \beta ^2 f'_0-f_0=0\,,
\end{equation}
where $f'_0=f'(R_0)$, $f_0=f(R_0)$ and $R_0=12\beta^{2}$. The \rf{dSCond}  also implies that not all the $f(R)$ theories of gravity admit de Sitters solutions, consistently with what one finds in \cite{Carloni:2004kp,SanteGenDynSys}. 

Let us now consider a perturbation of this spacetime in which a fluid, constituted for instance by standard matter, is present. According to what has been said in the previous sections, this fluid will be described by first-order quantities. We will also assume that the fluid is actually barotropic in its rest frame i.e. its equation of state is $p^m=w \mu^m$. Choosing a set of observers comoving with it\footnote{Since the definition of the fluid flow $u_a$ is made at the level of the perturbed Universe this choice is legitimate. In addition to that, the de Sitter solution is frame invariant so any choice of frame in the background is equivalent.}, the perturbations equations turn out to be \cite{Carloni:2009gp}
\begin{subequations}\label{PertScadS}
\begin{flalign}
&\dot{\Delta}=-\Theta (w +1) \Delta\,, \\
&\dot{\mathcal{R}}=\Re\,,\label{eqRdS}\\
&\dot{\Re}=-\Theta \Re-\frac{1}{3}\left[\frac{ f'}{f''}-R\right]\mathcal{R}+\prjnab^2 \mathcal{R} +\frac{(1-3 w ) }{3 f''} \Delta \label{eqRedS}\,.
\end{flalign}
\end{subequations}
where, as usual, we have assumed $f''(R)\neq0$ i.e. we are excluding the GR case. Note that these equations cannot be obtained plugging the background we defined directly from \rf{PertSca}. This because the fact that the matter thermodynamical variables are of first order changes the structure of linearized 1+3 equations, which, in turn, leads to changes in the differential structure of the equations rather than only in the coefficients.  
 Performing the harmonic decomposition and substituting \rf{eqRdS} in \rf{eqRedS} we obtain \cite{Carloni:2009gp}
\begin{subequations}\label{PertScadSHarm}
\begin{flalign}
&\dot{\Delta}^{(k)}=-3 \beta  (w +1) \Delta^{(k)}\,,\label{PertDeldSHarm} \\
&\ddot{\mathcal{R}}^{(k)}+3 \beta
   \dot{\mathcal{R}}^{(k)}+\left[\frac{ e^{-2 t \beta } k
   ^2}{S_0^2}-12 \beta^2+\frac{2
   f'_0}{f''_0}\right]\mathcal{R}^{(k)} -\frac{(1-3 w ) \Delta^{(k)} }{3 f''_0}=0\,.
\end{flalign}
\end{subequations}
In this system the equation for $\Delta$ is scale invariant and, as expected, matter perturbations are exponentially suppressed with a time constant which depends on $w$ and the time constant of the de Sitter solution. The Ricci scalar perturbations, instead, are governed by a second-order equation which is forced by the matter term. 

In the long wavelength limit $k=0$ the above equations yield the general solutions
\begin{eqnarray}
&&\Delta=\Delta _0 \,e^{-3 t  \beta  (1+w )} =\Delta _0 \left(\frac{S}{S_0}\right)^{-3 (w  +1 )}, \\
&& \mathcal{R}=\mathcal{R}_{0,1}\,e^{-3 t \beta  (1+w )} + \mathcal{R}_{0,2}\, e^{t \alpha _+}+\mathcal{R}_{0,3}\,e^{t \alpha _-},
\end{eqnarray}
where
\begin{equation}
\alpha_\pm=-3 \beta\pm\sqrt{25 \beta ^2-\frac{4 f'_0}{3 f''_0}}\,.
\end{equation}
Here $\mathcal{R}_{0,i}$  and $\Delta _0$ are constants of integration and we have dropped the apex ``${(0)}$" to make the notation lighter. It is plain from this solution that, in a de Sitter background, standard matter is clearly made homogeneous, but this is not the case for the perturbation of the Ricci curvature. If one considers $\mathcal{R}$ as representing the scalar gravitational waves normally associated to the scalar degree of freedom of this type of theories, one can see that, depending on the form of the function $f$, this kind of perturbation is able to grow. In addition, if we imagine our $f(R)$-model to be an inflationary one, we can see that the analysis of the scalar waves would constitute a direct and purely classical test of the nature of the gravitational interaction, based on the gravitational wave relics of the inflationary era. 

The form of the exponents of the modes the solutions above for some popular models of $f(R)$-gravity \cite{shosho,Nojiri:2007as,Cognola:2007zu} are given in Table \ref{dSTab}.

It is easy now to derive the properties of the power spectrum for the matter perturbations. It is clear from \rf{PertDeldSHarm} that their spectrum it is scale invariant and remains constantly scale invariant at all times.  Comparing these results with the ones of the previous section one realizes that $f(R)$  de Sitter inflation is  very different from an $f(R)$ power law inflation.  This result is similar to what happens in GR when de Sitter and power law inflation  are compared: also in this case power law inflation is accompanied by a loss of power in the infrared part of the spectrum \cite{Lucchin:1984yf}. 

The fact that with a de Sitter background  the power spectrum of the theory is flat seems to be in contrast with the claim of the presence of characteristic signature made in the previous section. However one must remember that the reason why this happens is the very special properties of the de Sitter metric and  the fact that in our background there is no matter. This eliminates all the $k$ dependence from the equation of the matter perturbations giving rise to a scale invariant spectrum. Such result reminds us that, independently from the form of $f(R)$, a specific choice of the background can change deeply the system \rf{SysIOrdHarm} and conditions the appearance of our characteristic signature. Therefore one can conclude that, in the appearance of  the characteristic signature in the power spectrum the structure of the perturbation equations due to the background  is more important than the additional scales contained in the function $f(R)$. This result shows again how the choice of a correct background is crucial to make any reliable prediction in perturbation theory.

\begin{table}[htdp]
\caption{Some of the values of $\beta$  and the exponents of the modes of the scalar fluctuation solutions for various popular $f(R)$-gravity models in pure de Sitter backgrounds \cite{Carloni:2009gp}. For the more complex forms of $f(R)$ the implicit equations to be solved in order to find the parameters have been given. Of special interest are the models $f(R)=\frac{R^m+\chi}{1+\xi R^n}$ and their generalizations, which can provide a unique theoretical framework for early time inflation and late time acceleration \cite{Nojiri:2007as,Cognola:2007zu} (the first unified models of this type were proposed in \cite{Nojiri:2003ft,Capozziello:2006dj}).}\label{dSTab}
\begin{center}
\scriptsize{\begin{tabular}{cccc}
\hline\hline
 $f(R)$ & $\beta$ & $\alpha _\pm$
 \\\hline
$R+\chi R^n$ &  $\left(2^{2 n-1} 3^{n-1} \alpha -3^n 4^{n-1} n \chi \right)^{\frac{1}{2-2 n}}$ & $-3 \beta \pm A$
\\&&&\\
$\exp(q R)$ &  $\pm\frac{1}{3\sqrt{2q}}$ & $\pm\frac{2}{3}\sqrt{\frac{2}{q}}$
\\&&&\\
$\frac{R^m+\chi}{1+\xi R^n}$ &
$\frac{3^{n+1} 4^n n \beta ^{2 n} \left(12^m \xi  \beta ^{2 m}+1\right)-\left(3^m \left(3 4^m m+2^{2 m+1}\right) \xi
   \beta ^{2 m}+2\right) \left(12^n \beta ^{2 n}+\chi \right)}{12 \left(12^m \xi  \beta ^{2 m}+1\right)^2}=0$ &$-3 \beta \pm \sqrt{25\beta ^2+B }$
 \\&&&\\
$R+\chi+\frac{\chi }{\alpha  \left[(R \beta -1)^{2
   n+1}+1\right]+1}$& $\beta ^2-\frac{6 \beta ^2 \xi ^2 \chi ^2 (2 n+1) \left(12 \beta ^2 \xi -1\right)^{2 n}}{\left(2 \left(12
   \xi -1\right)^{2 n+1}+\xi  \chi +2\right)^2}+\frac{\chi}{6} -\frac{\xi  \chi ^2}{12 \left(12 \beta ^2
   \xi -1\right)^{2 n+1}+\xi  \chi +2}=0$& $-3 \beta \pm \sqrt{25 \beta ^2-D}$
 \\&&&\\\hline\\
\multicolumn4c{$A=\sqrt{\frac{25 n^2+7n-32}{3n (n-1)}}\left(2^{2 n+3}-3 n 4^n\right)^{\frac{1}{2-2 n}} \chi ^{\frac{1}{2-2 n}}$}\\\\\multicolumn4c{$B=\frac{16 \left(12^m \xi  \beta ^{2 m}+1\right) \left(12^n n \beta ^{2 n} \left(12^m \xi  \beta ^{2 m}+1\right)-12^m
   m \beta ^{2 m} \xi  \left(12^n \beta ^{2 n}+\chi \right)\right)}{12^m m \xi  \left(12^m (m+1) \xi  \beta ^{2
   m}-m+1\right) \left(12^n \beta ^{2 n}+\chi \right) \beta ^{2 m}+12^n (n-1) n \left(12^m \xi  \beta ^{2
   m}+1\right)^2 \beta ^{2 n}-2^{2 m+2 n+1} 3^{m+n} m n \xi  \left(12^m \xi  \beta ^{2 m}+1\right) \beta ^{2 (m+n)}}$}\\\\\multicolumn4c{$C=\frac{2 \left(24 n \beta ^2 \xi  \chi  \left(144 \xi  \beta ^4+1\right)^{-n-1}+1\right)
   \left(144 \xi  \beta ^4+1\right)^{n+2}}{3 n \xi  \left(144 (2 n+1) \beta ^4 \xi -1\right) \chi }$}\\\\\multicolumn4c{$D=\frac{\left(12 \beta ^2 \xi -1\right)^{1-2 n} \left(2 \left(12 \beta ^2 \xi -1\right)^{2 n+1}+\xi  \chi +2\right)^3}{3 (2 n+1) \xi ^3 \chi ^2 \left(2 (n+1) \left(12 \beta ^2 \xi -1\right)^{2
   n+1}-n (\xi  \chi +2)\right)}
   \left(1-\frac{2 (2 n+1) \xi ^2 \left(12 \beta ^2 \xi -1\right)^{2 n} \chi ^2}{\left(2 \left(12 \beta ^2 \xi
   -1\right)^{2 n+1}+\xi  \chi +2\right)^2}\right)$}\\ \\ \hline\hline
\end{tabular}}
\end{center}
\end{table}

\section{Discussion and Conclusions}
In this paper I have reviewed the construction of a covariant and gauge invariant formalism that allows the description the perturbations around any cosmological background for a generic $f(R)$ theory of gravity. Using this technique I have been able to derive, in a frame comoving with matter, a system of equations that describes the evolution of the  scalar perturbations which are traditionally associated with structure formation. The covariance of the formalism guarantees that we are able to write these equations in any other frame. Differently from the  case of GR, this system appears to be of order four rather two, and the coefficient of the equations depends on derivatives up to four of the scale factor. Also, in their harmonic decomposed form, the  equations present a non trivial dependence on the wavenumber $k$, so that perturbations of dust can depend on the scale in spite of the absence of fluid pressure.  Such features are originated by the additional Laplacians appearing in the equations which, in turn, are a direct consequence of the fourth order terms in the gravitational field equations. 

The application of these equations to some simple $f(R)$-theories helps to further clarify their properties. We have seen that the behavior of the perturbations in these theories can be very different from the one in GR. For example, in $R^n$-gravity long wavelength perturbations can grow in a power law inflationary scenario. Also we learned how important are the features of the background in the determination of the behavior of the perturbations. For example, form the analysis 
of the de Sitter backgrounds one learns that the specific $k$-structure of the equations can be changed by the use of backgrounds with special properties. Our results also help in the understanding of more general features of perturbations in $f(R)$-gravity. The comparison of the results in $R^n$ and $R+\alpha R^n$ on the same background, shows us that the presence of an additional universal constant in the action (and the associated lenghtscale), although introducing changes in the dynamics, has little influence on the power spectrum. In this sense the structure of the power spectrum, should preserve some basic characteristics, like the presence of at least three regimes of which two are scale invariant (but not necessarily at the same power). 

Can we speak then of ``characteristic signature" of $f(R)$-gravity? And in what sense? The information that we have discovered so far seems to show that the introduction of additional scales i.e.  additive terms in the function $f$ does not affect the basic features of the matter power spectrum, because they are only related to the $k$-structure of the perturbation equations. However, such structure can be modified by the specific features of the background. In this sense one can conclude  that a class of background which do not change $k$-structure of the equation are bound to generate  for any (non pathological) $f(R)$ theory spectra with similar features in terms of regimes and behavior at large and small scales. This  similarity is what we call a ``characteristic signature" of these models.

On a more observational  point of view the important consequence of this conclusion is that a deviation from the  scale invariance of the power spectrum can be interpreted as caused by non Einstenian gravitational interaction. Therefore the large scale structure surveys can be used as powerful tests for the gravitational interaction.

The same hold in the case of inflationary models based on $f(R)$ gravity. As in the case of GR, power law inflation is different from de Sitter inflation, and the detection of scalar gravitational waves, could not only be a footprint of non Einsteinian gravity at work, but also could allow us to distinguish between power law inflation and de Sitter one, and even give information on what form of $f(R)$ gravity is most likely to be at work. 

All these results look very promising. In spite of the many difference between $f(R)$-gravity and GR, the observable seem not completely in contrast with the observations. Of course the next step is now to try to refine the rough models we have used up to now to obtained better results. These will help shedding more light on this fascinating topic.

\appendix
\section*{Acknowledgments}
The author was funded by Generalitat de Catalunya through the Beatriu de Pinos contract 2007BP-B1 00136.

\end{document}